\documentclass[11pt]{article}

\pdfoutput=1

\usepackage{amsmath,amssymb,amsfonts,amscd,mathrsfs}
\usepackage{xcolor}
\definecolor{darkblue}{rgb}{0.1,0.1,.7}
\usepackage[colorlinks, linkcolor=darkblue, citecolor=darkblue, urlcolor=darkblue, linktocpage]{hyperref} 
\usepackage[square, comma, compress,numbers]{natbib}
\usepackage[]{graphicx}
\usepackage{geometry}
\usepackage[margin=10pt,font=small,labelfont=bf]{caption}
\usepackage{ifthen}
\usepackage{tikz}
\usepackage{subcaption}

\usepackage{dsfont}

\def\red{\color [rgb]{0.9,0.1,0.1}}

\newcommand{\reef}[1]{(\ref{#1})}

\def\eps{\epsilon}
\newcommand{\beq}{\begin{equation}} 
\newcommand{\eeq}{\end{equation}}
\def\del {\partial} 
\def\nn{\nonumber} 
\def\bZ {\mathbb{Z}} 
\def\bR {\mathbb{R}} 
 
\def\calO {{\cal O}}

\def\calM {{\cal M}} 
\def\calH {{\cal H}}

\def\calE {{\cal E}} 
\def\bZ {\mathbb{Z}} 
 
\def\bP {\mathbb{P}} 
\def\half{{\textstyle\frac 12}}

\def\ge{\geqslant}
\def\le{\leqslant}

\newcommand{\diffop}[2]{\ifthenelse{\equal{#2}{1}}{\frac{\mrm{d}}{\mrm{d} #1}}{\frac{\mrm{d}^#2}{\mrm{d} #1^#2}}}
\newcommand{\NO}[1]{{:\!#1\!:}}

\newcommand{\braket}[3]{\langle #1|#2|#3 \rangle}

\newcommand{\ket}[1]{|#1\rangle}
\newcommand{\bra}[1]{\langle #1|}

\newcommand{\mrm}[1]{{\mathrm #1}}

\usepackage[normalem]{ulem}

\def \nmax{n_{\rm max}}
\def \lat{{\#}}
\def\phys{{\rm ph}}
\def\ph{{\rm ph}}
\def\del{\partial}

\def \Emax{E_{\rm max}}
\def \Htr{{H_{\rm trunc}}}
\def \Hren{H_{\rm ren}}
\def \Er{{\calE_{\rm *}}}

\usepackage{accents}
\newlength{\dhatheight}

\numberwithin{equation}{section}
\begin{document}

\vspace*{-.6in} \thispagestyle{empty}
\begin{flushright}
CERN PH-TH/2014-254\\
\end{flushright}
\vspace{1cm} {\Large
\begin{center}
{\bf Hamiltonian Truncation Study of the $\phi^4$ Theory\\ in Two Dimensions}\\
\end{center}}
\vspace{1cm}
\begin{center}
{\bf Slava Rychkov$^{a,b,c}$,  Lorenzo G.~Vitale$^d$}\\[2cm] 
{
$^{a}$ CERN, Theory Division, Geneva, Switzerland\\
$^{b}$ Laboratoire de Physique Th\'{e}orique de l'\'{E}cole normale sup\'{e}rieure, Paris, France\\
$^{c}$ Facult\'{e} de Physique, Universit\'{e} Pierre et Marie Curie, Paris, France\\
$^d$ Institut de Th\'eorie des Ph\'enom\`enes Physiques, EPFL, CH-1015 Lausanne, Switzerland\\
}
\vspace{1cm}
\end{center}

\vspace{4mm}

\begin{abstract}
We defend the Fock-space Hamiltonian truncation method, which allows to calculate numerically the spectrum of strongly coupled quantum field theories, by putting them in a finite volume and imposing a UV cutoff. The accuracy of the method is improved via an analytic renormalization procedure inspired by the usual effective field theory. As an application, we study the two-dimensional $\phi^4$ theory for a wide range of couplings. The theory exhibits a quantum phase transition between the symmetry-preserving and symmetry-breaking phases. We extract quantitative predictions for the spectrum and the critical coupling and make contact with previous results from the literature. Future directions to further improve the accuracy of the method and enlarge its scope of applications are outlined.
 \end{abstract}
\vspace{.2in}
\vspace{.3in}
\hspace{0.7cm} December 2014

\newpage

{
\setlength{\parskip}{0.05in}
\renewcommand{\baselinestretch}{0.7}\normalsize
\tableofcontents
\renewcommand{\baselinestretch}{1.0}\normalsize
}

\setlength{\parskip}{0.1in}

\section{Introduction}

How do we extract predictions about a strongly coupled quantum field theory (QFT) which is not exactly solvable? The lattice would be one answer, but it's not the only one. Hamiltonian truncation techniques, which generalize the Rayleigh-Ritz method familiar from quantum mechanics, are a viable deterministic alternative to the lattice Monte Carlo simulations, at least for some theories. These techniques remain insufficiently explored, compared to the lattice, and their true range of applicability may be much wider than what is currently believed. There exist several incarnations of Hamiltonian truncation, some better known than others, differing by the choice of basis and of the quantization frame. For example, Discrete Light Cone Quantization (DLCQ) \cite{Brodsky:1997de} and Truncated Conformal Space Approach (TCSA) \cite{Yurov:1989yu} are two representatives of this family of methods.

Here we will be concerned with what is perhaps the simplest setting for the Hamiltonian truncation---the $\phi^4$ theory in two spacetime dimensions. Moreover, we will consider the most straightforward realization of the method---we will quantize at fixed time rather  than on the light cone, and use the Fock space basis for the Hilbert space rather than the abstruse conformal bases.\footnote{The use of a conformal basis in two dimensions requires compactifying the scalar field \cite{Coser:2014lla}, see the discussion in section \ref{sec:tcsa}.} We will expand the $\phi^4$ Hamiltonian into ladder operators, as on the first page of every QFT textbook. We will however take this Hamiltonian more seriously than in most textbooks. Namely, we will use it to extract non-perturbative predictions, rather than as a mere starting point for the perturbative calculations. Concretely, we will (1)~put the theory into a (large) finite volume, to make the spectrum discrete, (2)~truncate the Hilbert space to a finite dimensional subspace of low-energy states, and (3)~diagonalize the truncated Hamiltonian numerically.

In spite or perhaps because of its extreme simplicity, this concrete idea has so far received even less attention than its more sophisticated cousins mentioned above. The only prior works known to us are \cite{Lee:2000ac,Lee:2000xna}.\footnote{A more extensive description of this work can be found in \cite{Salwen:2002dx} and \cite{Windoloski}. Another paper \cite{Brooks:1983sb} studied the two-dimensional Yukawa model without scalar self-interaction.} Here, we will follow up on these early explorations with our own detailed study.

While the basic idea and the qualitative conclusions of our work will be similar to \cite{Lee:2000ac,Lee:2000xna}, our implementation contains several conceptual and technical novelties. In particular, we will pay special attention to the convergence rate of the method, and will develop analytical tools allowing to accelerate the convergence, improve the accuracy, and better understand the involved systematic errors. 

The advances reported in this paper, as well as the ongoing progress in developing the other variants of the Hamiltonian truncation \cite{Katz:2013qua,Katz:2014uoa,Chabysheva:2014rra}, \cite{Giokas:2011ix,Lencses:2014tba,Hogervorst:2014rta} make us hopeful that in a not too distant future these methods will turn into precision tools for studying strongly coupled QFTs. 

The structure of the paper is clear from the table of contents. 
In section \ref{sec:probl_meth} we present the problem and the basic methodology used to study the spectrum numerically. 

Section \ref{sec:renormalization} elucidates the ideas behind the renormalization procedure, its implementations adopted in the numerical study, and provides some tests of the analytical results. The reader afraid of the technicalities may skip it. Yet it is precisely this section which is the theoretical heart of the paper. 

Section \ref{sec:phi4} contains the main numerical application of the work, i.e.~the calculation of the spectrum of the two-dimensional $\phi^4$ theory. The dependence of the numerical results on the physical and unphysical parameters is analyzed carefully, and an estimate of the critical coupling is provided. Computations were performed using a {\tt python} code included with the {\tt arXiv} submission.

In section \ref{sec:priorwork} we compare our method to the existing ones in the literature. Most of these prior studies focused in particular on the critical coupling estimates.

We conclude in section \ref{sec:conclusions}. Appendix \ref{sec:speed} presents some technical details useful for the practical implementation of the procedure. 
Appendix \ref{sec:pert} provides the perturbative checks of our method, alongside a discussion of the Borel-summability of the model.

We would like to mention right away that this paper was developed in parallel with Ref.~\cite{Hogervorst:2014rta} published three months ago and devoted to the TCSA approach in $d>2$ dimensions. The concrete example treated in \cite{Hogervorst:2014rta} was the $\phi^4$ theory in $d=2.5$ dimensions, which has the same phase structure as the $d=2$ case studied here. The attentive reader will notice many similarities in section \ref{sec:phi4} regarding the physics discussion, and in section \ref{sec:renormalization} regarding the renormalization procedure. However, concerning the latter, there is also a difference of principle which will be stressed in section \ref{sec:compMat} below.

\section{The problem and the method}
\label{sec:probl_meth}

\subsection{Hamiltonian}

\label{sec:probl_meth1}
We will be studying the two-dimensional $\phi^4$ theory, defined by the following Euclidean action:
\begin{gather}
S= S_0+g\int d^2 x\, \NO{\phi^4}\,, \label{eq:action}\\
S_0= \frac{1}{2} \int d^2 x\, \NO{(\del\phi)^2 + m^2 \phi^2}\,.
\end{gather}
Here $:\,:$ denotes normal ordering. Normal ordering of the free massive scalar action $S_0$ simply means that we set to zero the ground state energy density ({\it in infinite flat space}, and {\it before adding the quartic perturbation}).
The quartic interaction term is then assumed normal-ordered with respect to the mass $m$ appearing in the free action. In perturbation theory this simply corresponds to forbidding the diagrams with lines beginning and ending inside the same quartic vertex. In terms of operators, this means that we are adding counterterms \cite{Coleman:1974bu}:
\beq
\label{eq:NOphi4}
\NO{\phi^4}=\phi^4 - 6 Z \phi^2 + 3 Z^2\,. 
\eeq
Here
\beq
Z=\int \frac{d^2k}{(2\pi)^2}\frac{1}{k^2+m^2} 
\eeq
is a logarithmically UV-divergent quantity.

Although absent in \reef{eq:action}, below we will also need to consider perturbations given by the normal-ordered $\phi^2$ operator:
\beq
\label{eq:NOphi2}
\NO{\phi^2}=\phi^2 - Z. 
\eeq

The above equations specify what we mean by the theory in infinite flat space, and also define the mass parameter $m$ and the quartic coupling $g$ in terms of which we will parametrize the theory. All physical quantities (such as particles masses and S-matrix elements) are then finite functions of $m$ and $g$. Also the change of the ground state energy density due to turning on the coupling $g$ is finite and observable in this theory. This change can be thought of as the contribution of the theory \reef{eq:action} to the cosmological constant.

Since both $m$ and $g$ are dimensionful, physics depends on their dimensionless ratio $\bar g=g/m^2$, while $m$ (or $g$) sets the overall mass scale. We will assume $g>0$ to have a stable vacuum. Both signs of $m^2$ are interesting, but in this paper we will only consider the case $m^2>0$. Notice that this does not mean that we will always be in the phase of preserved $\bZ_2$ symmetry $\phi\to-\phi$, since the mass parameter undergoes renormalization. In fact, as we will see below, for $m^2>0$ and $\bar g>\bar g_c=O(1)$ the theory finds itself in the phase where the $\bZ_2$ symmetry is spontaneously broken. This is a nonperturbative phenomenon. For $\bar g\ll1$, the fate of the $\bZ_2$ symmetry is of course determined by the sign of $m^2$.

In this paper we will study the above theory not in infinite space but on a cylinder of the form $S_L^1\times \bR$, where $S_L^1$ is the circle of length $L$ and $\bR$ will be thought of as Euclidean time. We will impose the periodic boundary conditions around the circle. We will describe the theory on this geometry in the Hamiltonian formalism, taking advantage of the fact that the finite volume spectrum is discrete. 

Now, what is the Hamiltonian which describes the theory \reef{eq:action} on $S_L^1\times \bR$? The correct answer to this question involves a subtlety, so let us proceed pedagogically. 

We first discuss the Hamiltonian which describes the free massive scalar. In canonical quantization, the field operator is expanded into modes:
\begin{equation}
\label{eq:modeexp}
\phi(x) = \sum_{k} \frac{1}{\sqrt{2L  \omega_k}} \left( a_k e^{i k x} + a_k^\dagger e^{-i k x}\right)\,,
\end{equation}
where the momenta $k$ take discrete values $k = {2 \pi n}/{L}$, $n\in \mathbb{Z}$, $\omega_k = \sqrt{m^2 + k^2}$, and the ladder operators satisfy the usual commutation relations:
\begin{equation}
 [a_k , a_{k'}] = 0, \quad [a_k , a^\dagger_{k'}] = \delta_{n n'}\,.
\end{equation}
The Hilbert space $\calH$ of the theory is the Fock space of these ladder operators, spanned by the states  
\beq
\ket{\psi}=\ket{k_1,\ldots,k_m}=N a^\dagger_{k_1} \dots a^\dagger_{k_m} | 0 \rangle\,,
\label{eq:state}
\eeq
where $N$ is the normalization factor to get a unit-normalized state. 
The free scalar Hamiltonian is then given by:
\beq
H_{\text{free}} = H_0 + E_0(L),\qquad H_0=\sum_k \omega_k a^\dagger_k  a_k \,.
\label{eq:H0}
\eeq
The only subtlety here is the c-number term $E_0(L)$. The point is that we want the oscillator part $H_0$ of the finite volume Hamiltonian to be normal-ordered. However, the normal ordering counterterm in infinite space and for finite $L$ is slightly different, and $E_0(L)$ compensates for this mismatch. It is nothing but the Casimir energy of the scalar field, and is given by (see \cite{Bordag:2009zzd}):
\beq
\label{eq:E0L}
E_0(L)=-\frac{1}{\pi L}\int_0^\infty dx \frac{x^2}{\sqrt{m^2L^2+x^2}} \frac{1}{e^{\sqrt{m^2L^2+x^2}}-1}\, .
\eeq
This expression can be derived in many equivalent ways. One method is to regulate the difference of the zero-point energies:
\beq
\sum_n  \omega_{k_n}/2 - L \int_{-\infty}^{+\infty} \frac{dk}{2\pi} \omega_k/2 \, .
\eeq
Another method is to compute the partition function of the theory on the torus $S_{L_1}^1\times S_{L_2}^1$, which can be done from the path integral formulation of the theory. The partition function defined in this way enjoys the property of modular invariance. This method naturally produces a term in the free energy of the form $(2\pi L_2)\times E_0(L_1)$.

We next discuss the finite-volume Hamiltonian for the interacting theory. It will have the form:
\begin{gather}
\label{eq:partialham}
H = E_0(L)+ H_0+ g V_4+\ldots\,,\\
V_4=\int_0^L dx\,\NO{\phi^4}_L\,. 
\end{gather}
The normal ordering here is defined on the circle of length $L$ in the Hamiltonian sense, just putting all creation operators to the left. Thus:
\begin{multline}
V_4 = g L \sum_{k_1+k_2+k_3+k_4=0} \frac{1}{\prod \sqrt{2 L \omega_i}}
\Big[ a_{k_1}a_{k_2}a_{k_3}a_{k_4} +
4 a^\dagger_{-k_1}a_{k_2}a_{k_3}a_{k_4} \\ +
6 a^\dagger_{-k_1}a^\dagger_{-k_2}a_{k_3}a_{k_4} 
 +4 a^\dagger_{-k_1}a^\dagger_{-k_2}a^\dagger_{-k_3}a_{k_4} 
+ a^\dagger_{-k_1}a^\dagger_{-k_2}a^\dagger_{-k_3}a^\dagger_{-k_4} 
\Big]\,.
\end{multline}
The origin of the $\ldots$ terms in \reef{eq:partialham} lies again in the fact that the normal-ordering counterterms added when defining $V$, 
\beq
\NO{\phi^4}_L=\phi^4- 6 Z_L \phi^2 + 3 Z_L^2\,,\,\qquad Z_L=\sum_n \frac{1}{2L \omega_{k_n}}\,,
\eeq
are not exactly the same as in the infinite space definition \reef{eq:NOphi4}. The difference is
\beq
\NO{\phi^4}-\NO{\phi^4}_L=-6(Z-Z_L)\phi^2 +3(Z^2-Z_L^2)=6(Z_L-Z)\NO{\phi^2}_L +3(Z_L-Z)^2\,,
\eeq
where in the second equality we used $\phi^2=\NO{\phi^2}_L+Z_L$.

To compute $Z_L-Z$ we rewrite $Z$ in the form adapted to the Hamiltonian quantization:
\beq
Z=\int \frac{dk}{4\pi}\frac{1}{\sqrt{k^2+m^2}} \,.
\eeq
The difference $Z_L-Z$ is finite and is readily calculated using the Abel-Plana formula:
\beq
z(L)\equiv Z_L-Z=\frac{1}{\pi }\int_0^\infty \frac{dx}{\sqrt{m^2L^2+x^2}}\frac{1}{e^{\sqrt{m^2L^2+x^2}}-1}\, .
\eeq
This allows us to complete the $\ldots$ terms in \reef{eq:partialham}. Thus, the Hamiltonian on a circle of finite length $L$ corresponding to the infinite space theory \reef{eq:NOphi4} is given by:
\begin{gather}
\label{eq:ham}
H = H_0+ g[V_4+ 6 z(L) V_2]+ [E_0(L)+3 z(L)^2 g L],\\
V_2=\int_0^L dx\,\NO{\phi^2}_L=\sum_k \frac{1}{2 \omega_k}( a_k a_{-k} + a^\dagger_k a^\dagger_{-k} +2 a^\dagger_k a_k)\,. 
\end{gather}

We see that the Hamiltonian \reef{eq:ham} differs from the ``naive" Hamiltonian
\begin{equation}
\label{eq:hamiltonian}
H = H_0+ V\,,\quad V=g V_4 \,
\end{equation}
by ``correction terms", proportional to $E_0(L)$ and $z(L)$. The presence of these terms is conceptually important. 
They would be also straightforward to include into numerical analysis, for any $L$. However, in this paper we will be focussing on the case $L m \gg1 $. In this regime the corrections due to $E_0(L)$ and $z(L)$ are exponentially suppressed, and their numerical impact is negligible. For this reason, and to simplify the discussion, we will omit the exponentially suppressed corrections. With this proviso, from now on we will use the ``naive" Hamiltonian \reef{eq:hamiltonian}.

\subsection{Truncation}
\label{sec:probl_meth2}

We next explain the truncation method. We will work in the Hilbert space $\calH$ spanned by the free massive scalar states. The Hamiltonian $H$ acts in this space, and the problem is to diagonalize it. We thus use the free massive scalar states as a basis into which we expand the eigenstates of the interacting theory. Let us think of the Hamiltonian as an infinite matrix $H_{ij}$ where $i,j$ numbers the states in $\calH$:
\beq
H_{ij}=\braket{i}{H}{j}\,.
\eeq 
Notice that the states $\ket{i}$ as introduced above form an orthonormal basis of $\calH$. To find the spectrum of the theory in finite volume, we need to diagonalize the matrix $H_{ij}$. This diagonalization can be done separately in sectors having fixed quantum numbers corresponding to the operators commuting with the Hamiltonian.

The first such quantum number is the momentum: $[P,H]=0$. In this paper we will be working in the sector of states of vanishing total momentum: 
\beq
P=k_{1} + \dots + k_{m}=0\,.
\eeq 
In a large volume, the states of nonzero momentum should correspond to boosted zero-momentum states, and their energies should be related to zero-momentum energies by the Lorentz-invariant dispersion relation. It would be interesting to check this in future work. 

The second conserved quantum number is the spatial parity $\bP$, which acts $x\to -x$. It maps the state \reef{eq:state} into 
$\bP\ket{\psi}=\ket{-k_1,\ldots,-k_m}$. In this paper we will be working in the $\bP$-invariant sector,\footnote{The extension of our method to the $\bP$-odd sector is straightforward. We consider only the $\bP$-even sector, because we do not expect bound states with $\bP=-1$.} whose orthonormal basis consists of the states
\beq
\label{eq:symm}
\ket{\psi^{\rm sym}}=\beta(\psi)\bigl(\ket{\psi}+\bP\ket{\psi}\bigr)\,,
\eeq
where $\beta(\psi)$ is the normalization factor: 
 \beq
\beta(\psi)=
1/\sqrt{2}\ \text{  if  }\ \bP\ket{\psi}\ne \ket{\psi},\quad
1/2 \text{  otherwise}.
 \eeq
{\bf The restriction to the subspace $P=0, \bP=1$ will be tacitly assumed in all of the rest of the paper.}  
 
The final conserved quantum number is the already mentioned global $\bZ_2$ symmetry $\phi\to -\phi$ (the field parity). Its eigenvalue on the states \reef{eq:state} is $(-1)^m$. Below we will be considering both the $\bZ_2$-even and $\bZ_2$-odd sector. 

Each of the two sectors $\bZ_2=\pm 1$ still contains infinitely many states. We will thus have to truncate the Hilbert space. 
The truncation variable will be the $H_0$-eigenvalue:
\beq
E=\omega_{k_{1}} + \dots + \omega_{k_{m}}\,.
\eeq
We will truncate by considering all states of $E\le\Emax$. The parameter $\Emax$ should be thought of as a UV cutoff. The truncated Hilbert space is finite-dimensional, and the matrix $H_{ij}$ restricted to this space can be diagonalized numerically. This is what we will do.

In principle, one could imagine alternative truncation schemes.
For example, one can truncate in the maximal wavenumber $k_{\max}$. Such a truncation would be closer to the usual way one implements the UV cutoff in field theory. By itself, however, it does not render the Hilbert space finite-dimensional. One could also think of truncating in the total occupation number of the state, or in the individual occupation numbers per oscillator, and so on. Our initial exploration of such subsidiary cutoffs did not produce any dramatic gains in the performance of the method. In the end we decided to stick to the cutoff in $E$. As we will see in the next section, this cutoff allows for a natural implementation of the renormalization of the Hamiltonian, necessary to improve the convergence of the method. In the future it may be interesting to return to the other cutoffs, and explore them more systematically.

\section{UV cutoff dependence and renormalization}
\label{sec:renormalization}

\subsection{General remarks}

It is not difficult to write a code which computes the $H_{ij}$ matrix restricted to the $E\le \Emax$ subspace\footnote{See appendix \ref{sec:speed} for some tricks speeding up this computation.} and diagonalizes it.
The results of these numerical calculations will be discussed below. As we will see, as the UV cutoff $\Emax$ is increased, the energy levels computed using the truncated Hilbert space (`truncated energy levels') tend to some finite limits. These limits should be naturally identified with the exact energy levels. An interesting theoretical question then arises: \emph{what is the convergence rate of the method}? There is also a related practical question: \emph{how can the convergence be improved}? These questions will be discussed in this section.

By calculating the truncated energy levels we are discarding the contribution to the low-energy physics coming from the high energy states of the Hilbert space. Since the UV divergences have been already taken care of, this contribution is power-suppressed and goes to $0$ as the cutoff is increased.
In the standard Wilsonian approach to the renormalization group, by integrating out high-momentum (or short-distance) degrees of freedom one gets a flow in the space of Hamiltonians, along which the same physics is described in terms of low-momentum degrees of freedom with renormalized couplings. We would like to apply the same philosophy to our case, although we may expect some differences, because our cutoff prescription---cutting off in $E$---is different from the ones normally used in field theory. First of all, it breaks the Lorentz invariance. Second, the fact that we truncate in the \emph{total} energy of the state, rather than in that of its individual constituents, renders our cutoff effectively non-local. Thus, we should be prepared to see non-local as well as Lorentz-violating operators generated by the flow. We will see, however, that to leading order it will be sufficient to renormalize a few local operators in the Hamiltonian. It will be possible to do this computation in perturbation theory, since the potential we add to the free Hamiltonian is a relevant deformation and becomes less important in the UV. The dimensionless parameter which sets the convergence of the truncated energy levels  and the asymptotic magnitude of the counterterms will be ${g}/{\Emax^2}$. All these considerations will be made concrete in the following.

We start our analysis from the exact eigenvalue equation:
\beq
H.c=\calE c\,,
\label{eq:ex0}
\eeq
where $c$ is an infinite-dimensional vector living in the full Hilbert space $\calH$. Here and below, we use curly $\calE$ to denote energy levels of the interacting theory, while $E$ will be used to denote free scalar energy levels.

In our methodology the Hilbert space is divided in two subspaces: 
\begin{equation}
\mathcal{H} = \mathcal{H}_l \oplus \mathcal{H}_h\,,
\end{equation}
where $ \mathcal{H}_l$ is the low-energy sector of the Hilbert space, treated numerically, while $ \mathcal{H}_h$ is spanned by an infinite number of discarded high-energy states.
So we have $c=(c_l,c_h)^t$, and Eq.~\reef{eq:ex0} takes the following form in components:
\beq
H_{ll}.c_l+H_{lh}.c_h=\calE c_l\,,\qquad H_{hl}.c_l+H_{hh}.c_h=\calE c_h\,.
\label{eq:excomp}
\eeq
Here we denoted
\beq
H_{\alpha\beta}\equiv P_\alpha H P_\beta, 
\eeq
where $P_\alpha$ ($\alpha=l,h$) is the orthogonal projector on $\mathcal{H}_\alpha$.

Using the second equation in \reef{eq:excomp} to eliminate $c_h$ from the first one, we obtain:
\beq
\label{eq:ex}
[H_{ll}- H_{lh}.(H_{hh}-\calE)^{-1}.H_{hl}].c_l = \calE c_l\,,
\eeq 
or, equivalently,
\begin{gather}
\label{eq:ex1}
[H_{\rm trunc}+\Delta H].c_l = \calE c_l\,,\\
 \Delta H = - V_{lh}.(H_0+V_{hh}-\calE)^{-1}.V_{hl}\,.
\end{gather} 
This equation is very important. Notice that $H_{ll}\equiv H_{\rm trunc} $ is nothing but the Hamiltonian truncated to the low-energy Hilbert space. Notice furthermore that the mixing between the high and low-energy states is due only to $V$, since $H_0$ is diagonal. 

Eq.~\reef{eq:ex1} is \emph{exact}, yet it resembles the truncated eigenvalue equation, with a correction $\Delta H$. This equation will be a very convenient starting point to answer the two questions posed at the beginning of this section.
 
We will now start making approximations. First, we expand $\Delta H$ in $V_{hh}$ and keep only the zeroth term
\beq
 \label{eq:dhappr}
 \Delta H = - V_{lh}.(H_0-\calE)^{-1}.V_{hl}+\ldots
 \eeq
By dimensional reasons, we expect that the next term in the expansion, 
\beq
 \label{eq:nlo}
V_{lh}.(H_0-\calE)^{-1}.V_{hh}.(H_0-\calE)^{-1}.V_{hl},
\eeq 
will be suppressed with respect to the one we keep by $g/{\Emax^2}$. It will be very interesting to include this term in future work, and we will comment below about how this can be done. 
  
Equation \reef{eq:dhappr} defines $\Delta H$ as an operator on $\calH_l$. The definition depends on the eigenvalue $\calE$ that we are trying to compute. This subtlety will be dealt with below, while for the moment let us replace $\calE$ by some reference energy $\Er$. Even then, the definition seems impractical since it involves a sum over infinitely many states in $\calH_h$. Indeed, the matrix elements of $\Delta H$ according to this definition are given by:
\begin{equation}
\label{eq:dhfull}
(\Delta H)_{ij} =- \sum_{k:E_k>\Emax} \frac{ V_{ik} V_{kj}}{E_k-\Er}\,.
\end{equation}
Fortunately, in the next section we will give a simplified approximate expression for $\Delta H$ not involving infinite sums.
As we will see, to leading order $\Delta H$ will be approximated by a sum of local terms:
\begin{gather}
\label{eq:Vn}
\Delta H \approx \sum_N \kappa_N V_N , \qquad V_N= \int_0^L d x\, \NO{\phi(x)^N}\,.
\end{gather}
To this leading order, adding $\Delta H$ to $H_{\rm trunc}$ results in simply renormalizing the local couplings. As we will see, a more accurate expression for $\Delta H$ contains subleading corrections, which in general cannot be expressed as integrals of local operators. The appearance of these nonlocal corrections is due to the above-mentioned fact that truncating in total energy is not a fully local way of regulating the theory. 

\subsection{Computation of $\Delta H$}
\label{sec:deltah_comp}

Consider then the matrix elements \reef{eq:dhfull} of $\Delta H$ for $i,j$ in the truncated basis. We will write them in the form
\begin{gather}
\label{eq:deltah}
(\Delta H)_{ij} =- \int_{E_{\max}}^{\infty} d E\, \frac{ M(E)_{ij}}{E-\Er}\, ,\\
\label{eq:deltah1}
M(E)_{ij}\,dE \equiv \sum_{k:E \le E_k < E+ dE} V_{ik}V_{kj}\,.
\end{gather}
We are interested in the large-$E$ asymptotics for $M(E)_{ij}$. Of course, for finite $L$ the energy levels are discrete and this function should be properly thought of as a distribution (a sum of delta-functions). However, since the high-energy spectrum is dense, the fluctuations due to discreteness will tend to average out when integrating in $E$.
Below we will find a continuous approximation for $M(E)_{ij}$, valid on average. Such an approximation will be good enough for computing the integral in \reef{eq:deltah} with reasonable accuracy. A small loss of accuracy will occur because of the sharp cutoff at $E=\Emax$; this will be discussed below in sections \ref{sec:spectrumvsL} and \ref{sec:Emaxdep}.

Our calculation of $M(E)_{ij}$ will follow the method introduced in \cite{Hogervorst:2014rta}, section 5.3. It will be based on the fact that the same quantity appears also in the following matrix element:
\begin{equation}
\label{eq:ctau}
C(\tau)_{ij}= \bra{i} V(\tau/2) V(-\tau/2)\ket{j} = \int_0^\infty dE\, e^{-\left[ E -\left(E_i+E_j\right)/2\right] \tau} M(E)_{ij}\,,
\end{equation}
where we inserted a completeness relation in the second step. A word about notation: the Euclidean time dependence of various operators is always meant in the interaction representation, e.g.
\beq
V(\tau)=e^{H_0\tau} V e^{-H_0\tau}\,.
\eeq
If the time dependence is not shown, it means that the operator is taken at $\tau=0$.

Eq.~(\ref{eq:ctau}) says that $C(\tau)$ is basically the Laplace transform of $M(E)$. The leading non-analytic part of $C(\tau)$ for $\tau \to 0$ will come from the leading piece of $M(E)$ as $E \to \infty$. Our method will proceed by first extracting the leading non-analytic part of $C(\tau)$, and then taking its inverse Laplace transform to get at $M(E)$.

We will present the computation for a general case when the potential contains both $:\phi^2:$ and $:\phi^4:$ terms:
\beq
\label{eq:Vgen}
V= g_2 V_2+g_4 V_4 \,.
\eeq
Our Hamiltonian \reef{eq:hamiltonian} has $g_2=0$, $g_4=g$. Turning on $g_2\ne0$ corresponds to an extra contribution to the mass. Having this coupling will be useful for a check of the formalism in section \ref{sec:phi2test} below. 

We have 
\begin{equation}
\label{eq:ctau1}
C(\tau) =\sum g_n g_m \int_0^L d x \int_{-L/2}^{L/2} d z \NO{\phi(x+z, \tau/2)^{n}}\NO{\phi(x,-\tau/2)^{m}} \,,
\end{equation}
where we used periodicity and invariance under spatial translations. The non-analyticity of $C(\tau)$ for $\tau \to 0$ comes from the integration region where the product of two local operators is singular, i.e.~when they are inserted at near-coinciding points. Let us focus on one term in the sum, and rewrite it using Wick's theorem as:
\beq
g_n g_m \int_0^L d x \int_{-L/2}^{L/2} d z \sum_{0\le k\le \min(n,m)} f_{nm,n+m-2k}\, G_L(z,\tau)^k\NO{\phi(x+z, \tau/2)^{n-k}\phi(x,-\tau/2)^{m-k}} \,.
\label{eq:wick}
\eeq
Here $G_L(z,\tau)$ is the two-point function of $\phi$ in the free theory on the circle of length $L$. The $f$'s are integer combinatorial factors (operator product expansion coefficients):
\beq
f_{nm,n+m-2k}=\binom{n}{k} \binom{m}{k} k!\,.
\eeq
In \reef{eq:wick}, the leading non-analytic behavior as $\tau\to0$ will come from the propagator powers $G_L(z,\tau)^k$. The remaining normal-ordered operators can be Taylor expanded in $z$, $\tau$:
\beq
\label{eq:leadingOPE}
g_n g_m \int_0^L d x \int_{-L/2}^{L/2} d z \sum_{0\le k\le \min(n,m)} f_{nm,n+m-2k}\, G_L(z,\tau)^k\,[\NO{\phi(x)^{n+m-2k}}+O(\tau^2,z^2)]\,.
\eeq
The terms $O(z)$ are not shown because they will vanish upon integration. The terms $O(\tau^2,z^2)$ will produce a subleading singularity as $\tau\to 0$. The corresponding contributions to $M(E)$ will be suppressed by $m^2/\Emax^2$ compared to the leading ones. In this work these subleading contributions will be neglected, but it will be interesting and important to include them in the future.\footnote{The subleading contributions will give rise to new, derivative, operators in the Hamiltonian. Since our regulator breaks Lorentz invariance, the derivatives in $\tau$ and $z$ are not going to enter symmetrically in these subleading terms.}

Eq.~\reef{eq:leadingOPE} means that at leading order the correction Hamiltonian $\Delta H$ will contain terms of the form \reef{eq:Vn} with $N=n+m-2k$. To find the couplings $\kappa_N$, we need to evaluate the non-analytic part of the following quantities:
\begin{equation}
\label{eq:Iktau}
I_k(\tau) \equiv \int_{-L/2}^{L/2} d z \,  G_L(z,\tau)^k,\quad k=0,1,2,3,4\,.
\end{equation}
As we will see below, for $k=0,1$ the $\tau\to 0$ behavior will be analytic (for $k=0$ this is a triviality). This implies that only $N=0,2,4$ terms will be generated in \reef{eq:Vn}. 

To evaluate \reef{eq:Iktau}, we will need a few well-known facts about $G_L(z,\tau)$. In the infinite volume limit $L\to\infty$ the rotation invariance is restored, and the two-point function is a modified Bessel function of the second kind, depending on the distance $\rho=\sqrt{z^2+\tau^2}$:
\begin{equation}
\label{eq:prop}
G(\rho) = \frac{1}{2 \pi} K_0(m \rho)\qquad(L=\infty)\,.
\end{equation}
It has a logarithmic short distance behavior and decays exponentially at long distances:\footnote{$\gamma$ is Euler's constant.}
\begin{equation}
\label{eq:asym2pf}
G(\rho) \approx \begin{cases}
\displaystyle - \frac{1}{2 \pi} \log\left(\frac{e^{\gamma}}{2}  m\rho\right)[1 + O(m^2 \rho^2)]\,,  & \rho \ll 1/m\,,\\
\exp(-m\rho)/(2\sqrt{2\pi m \rho})\,, & \rho\gg 1/m\,.
\end{cases}
\end{equation}
For a finite $L$, the two-point function is obtained from the $L=\infty$ case via periodization:
\begin{equation}
G_L(z,\tau)=\sum_{n\in \bZ} G(\sqrt{(z+n L)^2+\tau^2})\,.
\end{equation}
The periodization corrections are exponentially small for $Lm\gg 1$. In our work, this condition will be always satisfied, and so we will use $G$ in place of $G_L$.\footnote{The induced error can be estimated by approximating $G_L(z,\tau)\approx G(\rho)+2G(L)$ for small $\rho$. This implies a shift $\Delta I_k(\tau)\approx \alpha I_{k-1}(\tau)$, $\alpha=2 k G(L)$. For $k=4$ and $L=4/m$ ($L=6/m$) the coefficient $\alpha=0.01 (0.002)$.} This is consistent with having neglected the exponentially suppressed $E_0(L)$ and $z(L)$ terms when passing from \reef{eq:ham} to \reef{eq:hamiltonian}.

So we will replace $G_L$ by $G(\rho)$ in \reef{eq:Iktau}. The non-analytic behavior of the integral comes from the small $z$ region, where the short-distance logarithmic asymptotic \reef{eq:asym2pf} is applicable. To regulate spurious IR divergences, it's convenient to calculate the first derivative with respect to $\tau$:
\begin{equation}
I_k'(\tau) = k \int_{-\infty}^\infty d z \,  (dG/d\rho) G(\rho)^{k-1} \frac{\tau}{\rho} 
\to k \left(-\frac{1}{2 \pi}\right)^k \int_{-\infty}^\infty dz \,  \left[ \log{\left(\frac{e^\gamma}{2} m \rho\right)}\right]^{k-1} \frac{\tau}{\rho^2}\,,
\end{equation}
where we also replaced $G$ by its short-distance asymptotics. The resulting integrals are convergent and readily evaluated:\footnote{{\tt Mathematica}'s {\tt Integrate} function sometimes gives wrong results for integrals of this type, so be careful.}
\begin{gather}
I_1'(\tau) = \mbox{const}\,,\nn
\\ I_2'(\tau) = \frac{1}{2 \pi} \log{m \tau} + \mbox{const}\,,\nn
\\ I_3'(\tau) = -\frac{3}{8 \pi^2} \left(\log{m\tau}\right)^2-\frac{3 \gamma}{4 \pi^2} \log{ m\tau }  + \mbox{const} \,,
\label{eq:I'}
\\ I_4'(\tau) = \frac{1}{4 \pi^3} \left(\log{m\tau}\right)^3+\frac{3 \gamma}{4 \pi^3} \left(\log{m\tau}\right)^2  +\frac{12 \gamma^2+\pi^2}{16 \pi^3} \log{m\tau}+ \mbox{const}\,\nn,
\end{gather}
modulo errors induced by using the short-distance asymptotics of $G$. These errors are suppressed by $O(m^2\tau^2)$. The corresponding corrections to $M(E)$ are suppressed by $m^2/\Emax^2$, and will be omitted. Also, as mentioned above, we see that $I_1'(\tau)$ is analytic. 

We now have to pass from the small-$\tau$ behavior to the large-$E$ asymptotics. Differentiating Eq.~(\ref{eq:ctau}) we have
\begin{equation}
\label{eq:ctauPrime}
C'(\tau)= \int_0^\infty d E\,  e^{-E_{i j}\tau} [-E_{i j} M(E)]\,,
\end{equation}
where we defined 
\beq
E_{i j} \equiv E - (E_i+E_j)/{2}\,.
\eeq
Thus from the inverse Laplace transforms of $I'_k(\tau)$ we should be able to determine the asymptotics of $-E_{i j} M(E)$. These inverse Laplace transforms are found from the following table of direct transforms:
\begin{gather}
\int_{\eps}^\infty dE\, e^{-E\tau}\frac 1E=-\log m\tau +\text{analytic}\,,\nn\\
\int_\eps^\infty dE\,e^{-E\tau} \frac {\log E/m}E=\frac 12(\log m\tau)^2+ \gamma\log m\tau +\text{analytic}\,,\\
\int_\eps^\infty dE\,e^{-E\tau} \frac {(\log E/m)^2}E=-\frac 13(\log m\tau)^3- \gamma(\log m\tau)^2-(\pi^2/6+\gamma^2)\log m\tau +\text{analytic}\,.\nn
\end{gather}
Since we are only interested in the large-$E$ asymptotics, the IR cutoff $\eps$ is not important---its value only influences the analytic parts.

Gathering everything, we obtain the following formula for the leading asymptotic behavior of $M(E)$:
\beq
\label{eq:me}
M(E)\sim  [g_4^2 \mu_{440}+ g_2^2 \mu_{220}] V_0 + [g_4^2 \mu_{442}+ g_2 g_4 \mu_{422}]V_2 + g_4^2 \mu_{444} V_4\Bigl|_{E\to E_{ij}}\,,
\eeq
where
\begin{gather}
\mu_{440}(E)=\frac{1}{E^2}\left\{\frac{18}{\pi^3}(\log E/m)^2-\frac{3}{2\pi}\right\},\qquad \mu_{220}(E)=\frac{1}{\pi E^2}\,,\nn\\
\mu_{442}(E)=\frac{72 \log E/m}{\pi^2 E^2}\,,\qquad \mu_{422}=\frac{12}{\pi E^2}\,,\qquad\mu_{444}(E)=\frac{36}{\pi E^2}\,.
\label{eq:me1}
\end{gather}
As the notation suggests, the $\mu$-functions in \reef{eq:me} are evaluated at $E=E_{ij}$.
This equation is the main result of this section. We subjected it to several tests, which we are going to describe below.

Before proceeding, let us comment on the evaluation of the next-to-leading term (\ref{eq:nlo}) in the renormalization procedure, which will be important in future developments of the method. From this term we will 
extract the $O(g^3/\Emax^4)$ contribution to the coefficients $\kappa_N$. This correction term is the most interesting of all $1/\Emax^4$ corrections, since it dominates in the limit $g\gg m^2$. Technically, we should generalize $C(\tau)$ and $M(E)$ in Eq.~\reef{eq:ctau} to functions of two variables ($\tau_{1,2}$ and $E_{1,2}$) and extract the leading non-analytic pieces for $\tau_{1,2} \to 0$. This calculation will involve Wick contractions among the operators in $C(\tau_1,\tau_2)$, the cyclic ones being the only nontrivial part.

We shall now move on to the tests of Eq. (\ref{eq:me1}).

\noindent{\bf Test 1}

\noindent Let us plug (\ref{eq:me}) into \reef{eq:deltah}, and do the integral neglecting the dependence on $\Er$ and ${(E_i+E_j)}/{2}$.\footnote{We stress that in numerical computations it will be important to retain these subleading corrections.} This gives $\Delta H$ of the form \reef{eq:Vn}, i.e.~as a sum of local counterterms with coefficients which are functions of $E_{\rm max}$. For example, the $g_4^2$ part is  given by (${\rm Log}\equiv \log\Emax/m$):
\beq
\label{eq:counterterm}
\Delta H \approx - \frac{g_4^2}{\Emax^2} \left\{ \left[\frac{9}{\pi^3} (\text{Log}^2 + \text{Log})+ \frac{3 (6 - \pi^2)}{4 \pi^3} \right] V_0  + 
\left(\frac{36}{\pi^2}\text{Log} + \frac{18}{\pi} \right) V_2 +
 \frac{18}{\pi} V_4
 \right\}\,.
\eeq
This expression was checked as follows. Working in infinite volume, we computed the order $g^2$ perturbative corrections to the vacuum energy, particle mass, and $2\to2$ scattering amplitude, imposing the cutoff $E\le \Emax$ on the intermediate state energy (thus working in the `old-fashioned' Hamiltonian perturbation theory formalism, rather than in terms of Feynman diagrams). We then checked that the leading $\Emax$ dependence of these results is precisely the one implied by \reef{eq:counterterm}. 
This way of arriving at \reef{eq:counterterm} is more laborious than the one given above, and we do not report the details.

\begin{figure}[t!]
\begin{center}
  \includegraphics[scale=0.66]{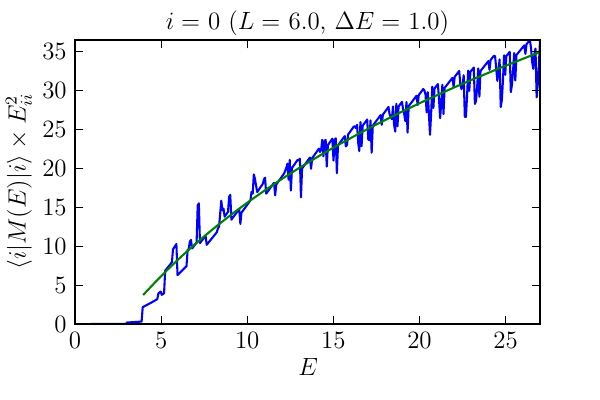}
  \includegraphics[scale=0.66]{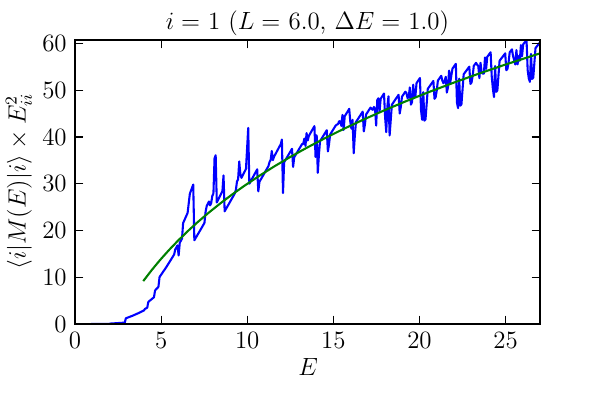}\\
  \includegraphics[scale=0.66]{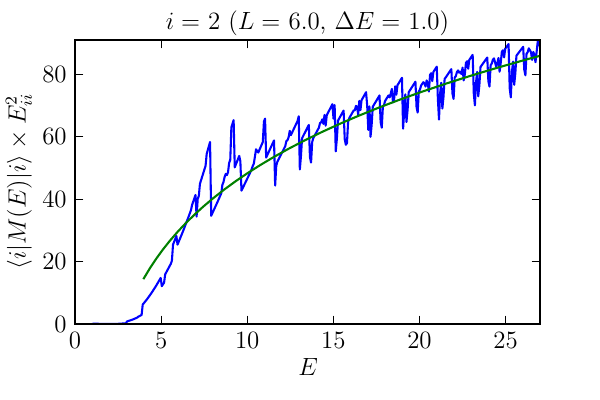}
  \end{center}
\caption{A test of the $M(E)$ asymptotics; see the text.}
\label{fig:me}
\end{figure}

\noindent{\bf Test 2}

\noindent A direct check of the asymptotics \reef{eq:me} can be done by comparing it with the actual value of $M(E)$ computed from its definition \reef{eq:deltah1}. One example is given in figure \ref{fig:me}, where we consider the diagonal matrix elements $\bra{i}M(E)\ket{i}$, $\ket{i}$ the state of $i$ particles at rest, $i=0,1,2$. We choose $m=1$, $L=6$, $g_2=0$ and $g_4=1$. The green smooth curves are the theoretical asymptotics from \reef{eq:me}. The blue irregular curves represent the moving average of $\bra{i}M(E)\ket{i}$ over the interval $[E-\Delta E,E+\Delta E)$ with $\Delta E=1$. To facilitate the comparison, both are plotted multiplied by $E_{ii}^2$. We see that the two curves agree quite well
on average.

A third test, involving the $g_2$ coupling, will be described in section \ref{sec:phi2test}.

\subsection{Renormalization procedures}
\label{sec:proc}

By ``renormalization", in a broad sense, we mean adding to the truncated Hamiltonian $\Htr$ extra terms designed to compensate for the truncation effects and reduce the $\Emax$ dependence of the results. In this section we will describe in detail the three renormalization prescriptions used in our numerical work.

Consider thus the Hamiltonian
\beq
\label{eq:g2g4}
H=H_0+V\,,\quad V=g_2 V_2 +g_4 V_4\,.
\eeq
In the main numerical studies in section \ref{sec:phi4} we will set $g_2=0$. The opposite case $g_4=0$, $g_2\ne 0$ will be considered in the check in section \ref{sec:phi2test}.

We are interested in the spectrum of $H$ on a circle of length $L$. Three approximations to this spectrum, in order of increasing accuracy, can be obtained as follows.

\noindent{{\bf 1.~Raw truncation} (marked `\emph{raw}' in plots)}

\noindent In this simplest approach, we are not performing any renormalization. The truncated Hamiltonian $\Htr$ is constructed by restricting $H$ to the subspace $\calH_l$ of the full Hilbert space, spanned by the states of energy $E\le \Emax$. The spectrum of $\Htr$ will be called the `raw spectrum'. According to Eqs.~\reef{eq:me}, \reef{eq:counterterm}, we expect that the raw spectrum approximates the exact spectrum with an error which scales as $1/\Emax^2$ (up to logarithms). 

\noindent{{\bf 2.~Local renormalization} (marked `\emph{ren.}' in plots)}

\noindent In this approach, we construct a correction Hamiltonian $\Delta H$ by the formula \reef{eq:deltah}. We use the asymptotics 
\reef{eq:me} for $M(E)$, in which we neglect $(E_i+E_j)/2$ with respect to $\Emax$. This gives a local $\Delta H$ of the form \reef{eq:Vn} with 
\begin{align}
\kappa_0&=-\int_{\Emax}^\infty\frac{dE}{E-\Er}[g_4^2 \mu_{440}(E)+ g_2^2 \mu_{220}(E)]\,,\nn\\
\kappa_2&=-\int_{\Emax}^\infty\frac{dE}{E-\Er}[g_4^2 \mu_{442}(E)+ g_2 g_4 \mu_{422}(E)]\,,\label{eq:kappas}\\
\kappa_4&=-\int_{\Emax}^\infty\frac{dE}{E-\Er}g_4^2 \mu_{444}(E)\,.\nn
\end{align}
The choice of the reference energy $\Er$ will be discussed shortly.
We then construct the `renormalized' Hamiltonian
\beq
\label{eq:Hren}
H_{\rm ren}=H_{\rm trunc}+\Delta H_{\rm loc},\qquad \Delta H_{\rm loc}\equiv\kappa_0 V_0+\kappa_2 V_2+ \kappa_4 V_4\,.
\eeq
Thus $\kappa_{2,4}$ correct the $g_{2,4}$ couplings, while $\kappa_0$ shifts the ground state energy density. Notice that the $\kappa$'s scale as $1/\Emax^2$ (up to logarithmic terms). 

The renormalized Hamlitonian acts in the same truncated Hilbert space $\calH_l$ as the truncated Hamiltonian $\Htr$. Its energy levels will be called the `renormalized spectrum'. This construction implements the first nontrivial approximation to the exact equation \reef{eq:ex1}. The local coupling renormalization accounts for the leading $1/\Emax^2$ error affecting the raw spectrum. Further corrections, discussed below, are suppressed by one more power of $\Emax$. So we expect that the renormalized spectrum approximates the exact spectrum with an error which scales as $1/\Emax^3$. 

Let us now discuss the reference energy $\Er$ in \reef{eq:kappas}. 
Recall that $\Er$ was introduced as a placeholder for the eigenstate energy $\calE$ in the definition \reef{eq:ex1} of $\Delta H$. Now, it's important to realize that the eigenstate energies do \emph{not} remain $O(1)$ in the limit of large $L$. The excitations \emph{above} the ground state, $\calE_I-\calE_0$,\footnote{We use small roman letters $i,j,\ldots$ to number states in the Fock space, which are eigenstates of $H_0$, and large letters $I,J,\ldots$ to number the eigenstates of the interacting Hamiltonian.} do stay $O(1)$, but the ground state energy itself grows linearly:
\beq
\calE_0\sim \Lambda L,\qquad L\to \infty\,.
\label{eq:calE0}
\eeq
Here $\Lambda$ is the interacting vacuum energy density (the cosmological constant), which is finite and observable in our theory.\footnote{Recall that the free vacuum energy density was set to zero by normal ordering the free scalar Hamiltonian.} 

We will therefore use the following recipe. We will choose $\Er$ close to, although not necessarily equal, the ground state energy of the theory. The precise choice will be specified when we present the numerical results. With this choice we compute the coupling renormalizations \reef{eq:kappas} and the renormalized spectrum. The differences $\calE_{I}-\Er$ will now be $O(1)$, and the error induced by this mismatch will truly be $1/\Emax$ suppressed. Moreover, even this error can be further corrected, as we discuss below.

We briefly mention here an alternative approach. One can insist that $\Er$ be adjusted, e.g.~iteratively, until it exactly equals the eigenvalue $\calE_I$ which comes out from diagonalizing $H_{\rm ren}$. This has to be done separately for each eigenstate, and so is rather expensive. We tried this method and found that it gives results in close agreement with those obtained from our simpler recipe for $\Er$, combined with the correction procedure described below.

\noindent{{\bf 3.~Local renormalization with a subleading correction} (marked `\emph{subl.}' in plots)}

\noindent We will now describe the third approach which improves on the previous one by taking into account not only the renormalization of the local couplings, but also the first subleading corrections due to the eigenstate energy and ${(E_i+E_j)}/{2}$. As explained above, these corrections can be considered smaller than the local ones by a further $O({1}/{\Emax})$ factor. They will take care of the mismatch between \reef{eq:deltah} and the local coupling renormalization. The corresponding correction Hamiltonian has the following matrix elements between the truncated Hilbert space states: 
\begin{equation}
\label{eq:subl}
[\Delta H_{\rm subl}(\calE)]_{i j} =(\lambda_{0})_{i j} (V_0)_{i j} + (\lambda_{2})_{i j} (V_2)_{i j} +  (\lambda_{4})_{i j} (V_4)_{i j}\,
\end{equation}
(no summation over the repeated indexes). The $(\lambda_N)_{i j}$ are the differences between the renormalization coefficients fully dependent on $(E_i+E_j)/{2}$ and $\calE$ and the local ones $\kappa_N$ defined in \reef{eq:kappas}:
\begin{align}
(\lambda_0)_{i j}&=-\int_{\Emax}^\infty\frac{dE}{E-\calE}[g_4^2 \mu_{440}(E_{i j})+ g_2^2 \mu_{220}(E_{i j})] -\kappa_0\,,\nn\\
(\lambda_2)_{i j}&=-\int_{\Emax}^\infty\frac{dE}{E-\calE}\,[g_4^2 \mu_{442}(E_{i j})+ g_2 g_4 \mu_{422}(E_{i j})]-\kappa_2\,,\label{eq:lambdas}\\
(\lambda_4)_{i j}&=-\int_{\Emax}^\infty\frac{dE}{E-\calE}\,g_4^2 \mu_{444}(E_{i j})-\kappa_4\,.\nn
\end{align}

There is a small technical subtlety in using the given expressions. For $(E_i+E_j)/2$ close to $\Emax$, the argument $E_{ij}$ of the $\mu$-functions is small in the part of the integration region close to $\Emax$. In this region it makes little sense to use \reef{eq:me1}, valid for large $E$. From figure  \ref{fig:me} we see that the asymptotics sets in roughly at $E\sim 5 m$. We therefore use the following prescription in evaluating \reef{eq:lambdas}: we use \reef{eq:me1} for $E_{ij}\ge 5m$, while we set $\mu$'s to zero below this threshold.

The full procedure is then as follows. We compute the local renormalized Hamiltonian \reef{eq:Hren} with the reference value $\Er$ fixed around the ground state energy. We diagonalize $\Hren$, determining the renormalized spectrum (in practice only a few lowest eigenvalues) and the corresponding eigenstates:
\begin{equation}
\Hren \ket{c_I} = \calE_{{\rm ren},I}\ket{c_I}
\end{equation}
Every eigenvalue is then corrected by adding \reef{eq:subl} at first order in perturbation theory:
\begin{equation}
\label{eq:subl1ord}
\calE_{{\rm subl},I}=\calE_{{\rm ren},I}+ \Delta \calE_{I}, \qquad \Delta \calE_{I} = \braket{c_I}{\Delta H_{\rm subl}(\calE_{{\rm ren},I})}{c_I}\,.
\end{equation}
From the computational point of view the evaluation of this correction can be considered inexpensive, since it scales as the square of the basis dimension, whereas the matrix diagonalization typically scales as its cube. The energy levels $\calE_{{\rm subl},I}$ will be called `renormalized subleading' or simply `subleading'. 

Second-order corrections can also be considered:
\begin{equation}
\Delta \calE_I^{(2)}= \sum_{J \ne I} 
\frac
{|\braket{c_I}
{\Delta H_{\rm subl}(\calE_{{\rm ren},I})}
{c_J}|^2}
{\calE_{{\rm ren},I}-\calE_{{\rm ren},J}}\,.
\end{equation}
These turn out to be negligible, except when there are two almost-degenerate eigenvalues.

\subsection{A test for the $\phi^2$ perturbation}
\label{sec:phi2test}

We will now perform a test of our method in a controlled situation when the exact answers are known.\footnote{This test is analogous to the one in \cite{Hogervorst:2014rta}, section 6.} Consider the theory described by the action (cf.~\reef{eq:action}):
\beq
S=S_0+g_2 \int d^2x :\phi^2:\,.
\eeq
The finite volume Hamiltonian corresponding to this problem has the form 
\beq
H=H_0+g_2 V_2+ C,\quad C=E_0(L)+ g_2 L z(L) \,.
\eeq
Just as in section \ref{sec:probl_meth1}, the extra constant term $C$ appears because of the difference in the normal ordering counterterms in the infinite space and on the circle. These terms are exponentially suppressed for $L m\gg 1$, but for the time being it will be instructive to keep them.

In full form, we have:
\beq
H=C+\sum_k \omega_k a_k^\dagger a_k + \frac{g_2}{2 \omega_k}( a_k a_{-k} + a^\dagger_k a^\dagger_{-k} +2 a^\dagger_k a_k)\,,\quad \omega_k=\omega_k(m)\,.
\label{eq:freemg2}
\eeq
We expect, of course, that this Hamiltonian corresponds to a free scalar of a mass
\beq
\label{eq:newmass}
\mu^2=m^2+2g_2\,.
\eeq
We will now use a Bogoliubov transformation to show this explicitly. The derivation is standard and is given here only for completeness. The transformation has the form
\beq
b_k = (\cosh \eta_k) a_k+(\sinh \eta_k) a_{-k}^\dagger
\eeq
with $\eta_k$ assumed real and depending only on $|k|$. The $b$'s then satisfy the same oscillator commutation relations as the $a$'s. We want to map \reef{eq:freemg2} onto
\beq
\sum_k \Omega_k b_k^\dagger b_k +\calE_0\,,\quad \Omega_k=\omega_k(\mu)\,.
\eeq
The conditions that the two Hamiltonians match take the form:
\beq
\Omega_k \cosh (2\eta_k)= \omega_k+g_2/\omega_k\,,\quad \Omega_k \sinh(2\eta_k) = g_2/\omega_k\,.
\eeq
This is indeed satisfied provided that
\beq
\Omega_k^2=\omega_k^2+2g_2\,,
\eeq
which proves the expression \reef{eq:newmass} for the new mass. The same derivation gives the value of the vacuum energy:
\beq
\label{eq:calE0g2}
\calE_0=C-\sum \Omega_k (\sinh \eta_k)^2=C+\frac 12\sum (\Omega_k-\omega_k- g_2/\omega_k)\,.
\eeq
Up to the constant $C$, the last expression can be intuitively understood \cite{Hogervorst:2014rta} by starting from the zero-point energy $\frac 12 \sum \Omega_k$ and subtracting the terms zeroth and first order in $g_2$. 

The series in \reef{eq:calE0g2} is convergent and can be summed using the Abel-Plana formula. We find that the constant $C$ is canceled, and the final result is given by
\beq
\label{eq:calE01}
\calE_0 = E_0(L,\mu)+\Lambda L,\qquad \Lambda = \frac{1}{8\pi}[\mu^2(1- \log \mu^2/m^2)-m^2]\,,
\eeq
where $E_0(L,\mu)$ is the Casimir energy of the free scalar field of mass $\mu$, given by \reef{eq:E0L} with $m\to\mu$.

The physical interpretation of \reef{eq:calE01} is clear. Apart from the usual Casimir energy term, we have an induced extensive vacuum energy, corresponding to a finite vacuum energy density $\Lambda$. Usually, when one studies the Casimir energy, the vacuum energy density in the infinite space limit is assumed to vanish. However, our situation here is different. We already finetuned to zero the vacuum energy density of the original, unperturbed, theory, i.e.~the one described by the action $S_0$. Once this is done, the vacuum energy density of the \emph{perturbed} theory becomes finite and observable. 

\begin{figure}[t!]
\begin{center}
  \includegraphics[scale=0.8]{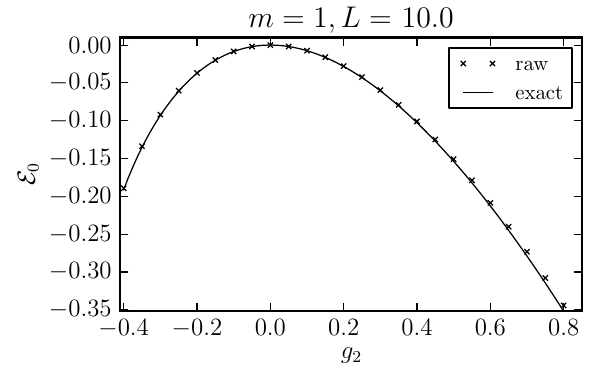}
  \includegraphics[scale=0.8]{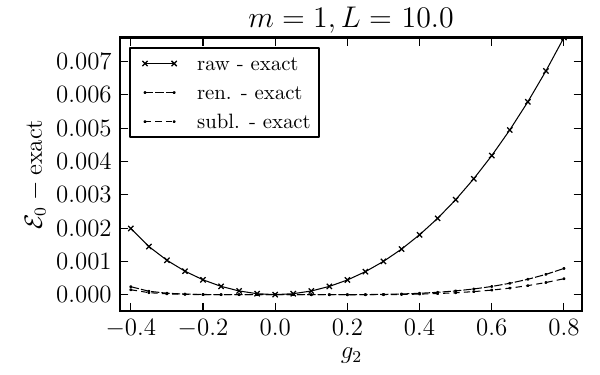}
  \end{center}
\caption{Exact and numerical ground state energy for the $\phi^2$ perturbation; see the text.}
\label{fig:phi2vac}
\end{figure}

We will now compare the above exact results with the numerical results obtained by using the Hamiltonian truncation. 
We will be considering the case $L m\gg 1$, which means that we will not be sensitive to the exponentially suppressed constant term $C$ in the initial Hamiltonian. We thus start directly from the Hamiltonian of the form \reef{eq:g2g4} with $g_4=0$, $g_2\ne 0$. We calculate its spectrum using the three procedures from section \ref{sec:proc}. In the shown plots we chose $m=1$, $L=10$, and varied $g_2$ from $-0.4$ to $0.8$.\footnote{The reference energy $\Er$ in \reef{eq:kappas} was set to the value of the ground state energy given by the raw truncation procedure.} For illustrative purposes numerics were done with a rather low cutoff $\Emax=12$, for which the truncated Hilbert space contains about 300 states. Figure \ref{fig:phi2vac} compares the ground state energy. In the left plot, the agreement between the raw and the exact result is already pretty good. The right plot shows the difference between the numerics and the exact value. We see that the renormalization greatly reduces the discrepancy over the raw procedure, and the results are made slightly better by including the subleading correction.

\begin{figure}[t!]
\begin{center}
  \includegraphics[scale=0.8]{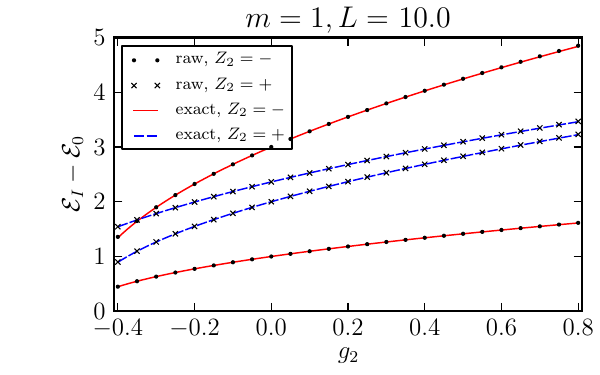}
  \includegraphics[scale=0.8]{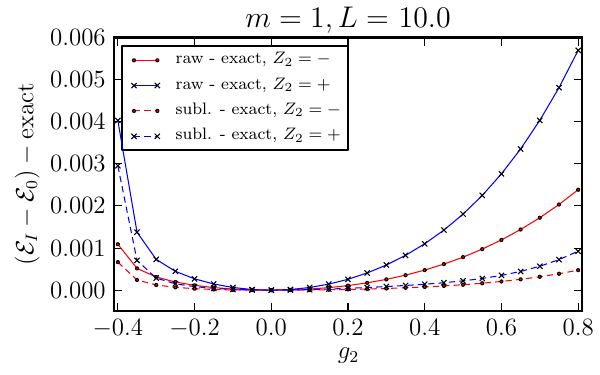}
  \end{center}
\caption{Exact and numerical spectra of excitations for the $\phi^2$ perturbation; see the text.}\label{fig:phi2spec}
\end{figure}

In figure  \ref{fig:phi2spec} we do the same comparison for the spectrum of excitations above the vacuum, $\calE_I-\calE_0$. In the left plot we pick the first two $\bZ_2$-odd states (one and three particles at rest), and the first two $\bZ_2$-even states (two particles at rest, and with one unit of momentum in the opposite directions). Already the raw spectrum agrees well with the exact values.
In the right plot we present the differences, focussing on the first two excited levels only (one even and one odd). Notice that for $g_4=0$ the difference between $\Hren$ and $\Htr$ is only in the vacuum energy coefficient $\kappa_0$, which shifts all eigenvalues in the same way. The first non-trivial corrections for the spectrum of excitations are therefore the subleading ones. The improvement over the raw results is significant. 

\subsection{Comparison to Ref.~\cite{Hogervorst:2014rta}}
\label{sec:compMat}

The reader will have noticed that our treatment of the UV cutoff dependence and renormalization is similar to Ref.~\cite{Hogervorst:2014rta}, sections 5 and 7.3. There is however a difference of principle that we will now explain.

Both in this work and in Ref.~\cite{Hogervorst:2014rta} the starting point for the
renormalization is Eq.~\reef{eq:ex1}. While Ref.~\cite{Hogervorst:2014rta} also presents this
equation, it then takes an alternative route, justifying the
renormalization procedure on the basis of the Hamiltonian perturbation
theory, see~\cite{Hogervorst:2014rta}, Eq. (5.8). This equation is then further subjected to
an RG improvement procedure in section 5.3 of \cite{Hogervorst:2014rta}, leading ultimately
to a result which differs from our Eq.~\reef{eq:ex1} only by some subleading
corrections.

Although the RG improvement introduced in~\cite{Hogervorst:2014rta} might be useful for
understanding the physical picture, it appears to be a detour that is
not strictly necessary for doing the computations. Eq.~\reef{eq:ex1} appears to provide the best starting point for the discussion of renormalization corrections.

A discussion on earlier approaches to analytic renormalization, in the context of TCSA, can be found section 5.4 of \cite{Hogervorst:2014rta}.

\section{Study of the $\phi^4$ theory}
\label{sec:phi4}

In the previous sections we have developed the method and tested it in the simple setting of the $\phi^2$ perturbation. We will now move on to the main task of this paper---to study the spectrum of the $\phi^4$ theory described by the Hamiltonian \reef{eq:hamiltonian}.

 The main physical parameter varied in our study will be the quartic coupling $g$. The physics depends on the dimensionless ratio $\bar g= g/m^2$, and we will work in the units where the mass term $m=1$. 
 
 The second parameter will be the size of the spatial circle $L$. This plays the role of the IR cutoff, to render the spectrum discrete. In practice one is usually interested in the infinite volume limit $L\to\infty$, and we will try to approach this limit. However, even a finite $L$ is physical, in the sense that the energy levels on the circle are well-defined physical observables. 
 
 The third parameter we will vary is the cutoff on the size of the Hilbert space $\Emax$ (the maximal energy of the free scalar Fock states included in the truncated Hilbert space). This parameter plays the role of the UV cutoff. It is unphysical. The continuum limit is recovered for $\Emax\to\infty$. 

We will typically present the results derived using the renormalization procedures both without (marked `ren.' in the plots) and with (marked `subl.') subleading corrections (see section \ref{sec:proc}). These procedures are expected to converge to the exact spectrum at the rate which goes as $1/\Emax^3$ and $1/\Emax^4$ (modulo logarithms). {\bf We take the difference between them as a rough idea of the current error of the method.}

\subsection{Varying $g$}

\begin{figure}[h!]
\begin{center}
 \includegraphics[scale=0.8]{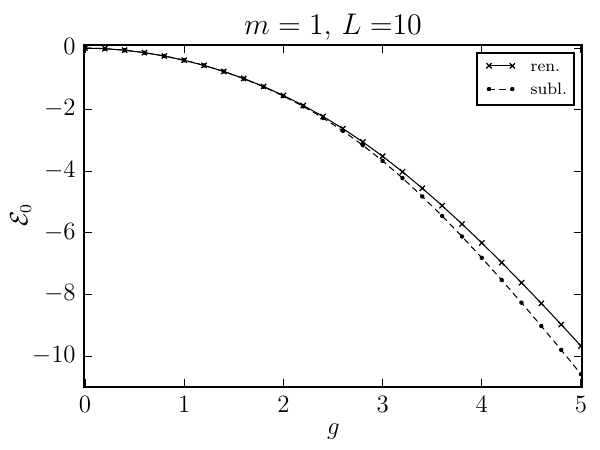} 
  \includegraphics[scale=0.8]{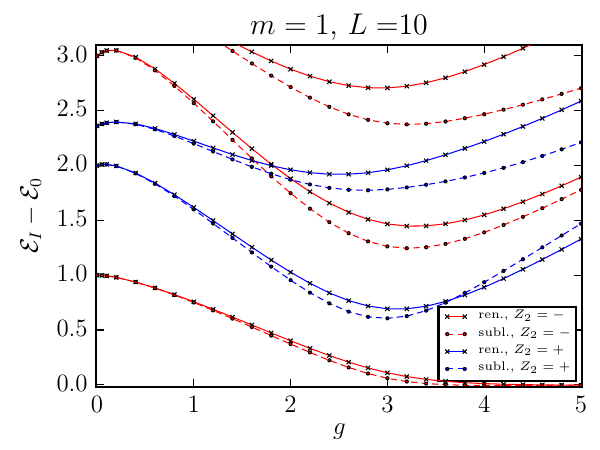}
   \end{center}
\caption{Numerical spectra as a function of $g$ for $m=1$, $L=10$; see the text.}
\label{fig:phi4}
\end{figure}

In figure \ref{fig:phi4} we present the ground state energy and the low energy spectrum of excitations for $g\le 5$. This extends well beyond the range $g\lesssim 0.5 - 1$ where perturbation theory is accurate (see appendix \ref{sec:pert}). In this plot we use a fixed value $L=10$, and choose the UV cutoff $\Emax=20$.\footnote{This corresponds to keeping 12870(12801) states in the even(odd) sector of the Hilbert space.} We use the two renormalization procedures explained in section \ref{sec:proc}.

The left plot shows the dependence of the ground state ($\equiv$ vacuum) energy on $g$. The vacuum is simply the state of the lowest energy, and it resides in the $\bZ_2$-even sector. There is not much structure in this plot, except for the fact that the vacuum energy is negative and grows in absolute value as $g$ is increased, becoming of the same order of magnitude as $\Emax$ for the largest $g$ considered here. This has a consequence for the renormalization procedure used in our study. Recall that in the local renormalization (the one marked `ren.') the 
coupling are renormalized using Eqs.~\reef{eq:kappas} which involve the reference energy $\Er$. {\bf Everywhere 
in this section we set $\Er$ to the value of the vacuum energy computed using raw truncation.} We already mentioned in section \ref{sec:proc} that since the vacuum energy may become large, the integrals in \reef{eq:kappas} have to be evaluated without expanding in $\Er$. We are fortunate here that the vacuum energy becomes large and \emph{negative}, and so the renormalization corrections become smaller if nonzero $\Er$ is taken into account.
A large and \emph{positive} vacuum energy would be a big problem for the performance of our method.\footnote{That the vacuum energy becomes negative both here and in section \ref{sec:phi2test} is probably more than just a coincidence. See the discussion in \cite{Hogervorst:2014rta}, note 21.}

The right plot shows the 5 lowest excitations above the vacuum, with the $\bZ_2=\pm$ excitations colored in blue(resp.~red). As we can see the first odd level becomes almost degenerate with the vacuum for $g \gtrsim 3$. This is a signal of the spontaneous $\bZ_2$-symmetry breaking. We therefore expect a second-order phase transition to occur at a critical point $g = g_c \approx 3$. For $g=g_c$, the theory should flow at large distances to a CFT. Since the $\phi^4$ theory is in the same universality class as the Ising model, we expect this IR CFT to be the minimal model $\calM_{4,3}$. We will analyze the region around $g = g_c$ in more detail below. For $g > g_c$ we are in the $\bZ_2$-broken phase. In this phase, the higher excitations should also be doubly degenerate in infinite volume. For a finite $L$ the exact degeneracy is lifted and becomes approximate. This degeneracy is not observed clearly in figure~\ref{fig:phi4}, probably because $L=10$ is not large enough.\footnote{The discussed phase diagram is the same as for the $\phi^4$ model in $d=2.5$ dimensions studied in \cite{Hogervorst:2014rta} using the TCSA. In that case it was possible to observe approximate degeneracy for the first and second excited states.}

In the region of small $g$, it is possible to validate the numerical results by comparing them to perturbation theory. In appendix \ref{sec:pert}, we do this comparison for the ground state energy and the mass of the lowest excitation. For small $g$, we find good agreement with the perturbative predictions computed through $O(g^3)$.

\begin{figure}[h!]
\begin{center}
 \includegraphics[scale=0.8]{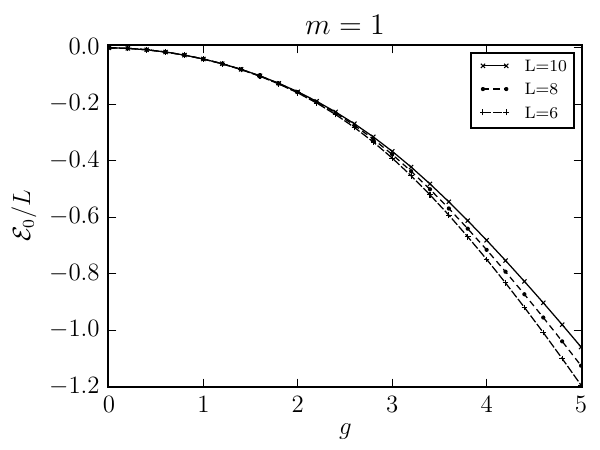} 
  \includegraphics[scale=0.8]{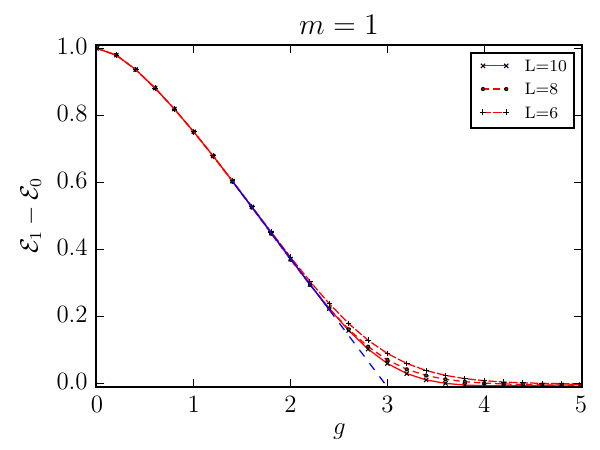}
  \end{center}
\caption{The vacuum energy (left) and the first odd excitation (right) determined numerically for $L=6,8,10$. The blue dashed line in the right plot is the fit to determine the critical coupling; see section \ref{sec:fixed}.}
\label{fig:phi4_multL}
\end{figure}

It is interesting to understand the sensitivity of the spectrum plot in figure \ref{fig:phi4} to the chosen value of $L=10$.
We therefore show in figure \ref{fig:phi4_multL} similar plots for $L$ equal to $6$, $8$, $10$ and $\Emax$ respectively equal to $34$, $26$ and $20$.\footnote{$\Emax$ is adjusted to have roughly the same size of the Hilbert space in all three cases. Smaller $L$ give larger energy spacings for the one-particle momentum excitations, and allow to go to larger $\Emax$.} To avoid clutter, only the results for the subleading renormalization (the third, most precise method in section \ref{sec:proc}) are presented. 

In the left plot we show the vacuum energy \emph{density} $\Lambda = \mathcal{E}_0/L$. For a sufficiently large $L$ this is supposed to become independent of $L$. We see that this constancy is verified with an excellent accuracy for $g\lesssim 2$. In this region we are in the massive phase, and the finite $L$ corrections are expected to be exponentially small (see section \ref{sec:spectrumvsL} below). The dependence on $L$ becomes more pronounced around $g=g_c$, which is as it should be because the mass gap goes to zero here. However, in the $\bZ_2$-broken phase the corrections remain significant, while theoretically they should become again exponentially suppressed. Therefore, for $g\gtrsim 3$, we are forced to interpret the variation with $L$ not as a physical effect but being due to finite $\Emax$ truncation effects. This is consistent with the significant difference between the results obtained with the two renormalization procedures in figure \ref{fig:phi4}.

In the right plot of figure \ref{fig:phi4_multL} we show the physical particle mass $m_\phys = \mathcal{E}_1-\mathcal{E}_0$. Once again, in the $\bZ_2$ unbroken massive phase there is hardly any dependence on $L$, while around $g=g_c$ there appears variation, which will be studied quantitatively in section \ref{sec:fixed} below.
This plot will also be used below to extract an estimate of $g_c$. 

Overall, the truncation effects seem to be too large for $g\gtrsim 3$ to allow precise quantitative claims about this range of couplings (apart from the fact that the $\bZ_2$ symmetry appears broken). Head-on treatment of that range would require a refinement of the method, by improving the renormalization procedure. An alternative way to access this region is to use the strong/weak coupling duality due to Chang \cite{Chang:1976ek}. In a companion work \cite{Rychkov:2015vap} we will both test this duality, and use it to study the $\bZ_2$-broken phase of the model.

\subsection{The critical point}
\label{sec:fixed}
We will now try to determine with some precision the critical coupling $g_c$, and study the lowest operator dimensions of the CFT at the phase transition. According to the standard renormalization group theory, for $g$ close to $g_c$ the physical mass  $m_\phys$ should behave as:
\begin{equation}
\label{eq:critExp}
m_\phys \sim C |g - g_c|^\nu, 
\end{equation}
where $C$ is a theory-dependent constant,\footnote{Which also depends on from which direction one approaches the fixed point.} and $\nu$ is a critical exponent, common for all theories in the Ising model universality class, and expressible via the dimension of the most relevant $\bZ_2$-even scalar operator, $\eps$, of the CFT: 
\beq
\nu = (2-\Delta_\epsilon)^{-1}\,.
\eeq

We used our numerical results obtained for $L=10$, $\Emax=20$ renormalized with subleading corrections (see figure \ref{fig:phi4_multL}) to perform the fit of $m_\phys\equiv \mathcal{E}_1 -\mathcal{E}_0$ to the formula \reef{eq:critExp}, replacing $\sim$ by $=$. 
Admittedly, our procedure is careless, since we do not take into account the corrections to scaling. We view the results which we will now present as preliminary; they should be validated by future studies as our method progresses. Another uncertainty concerns the range of $g$ chosen to do the fit. On the one hand, $g$ should be close to $g_c$, on the other hand right close to $g_c$ the spectrum is modified by finite size corrections. Looking at the right plot in figure \ref{fig:phi4_multL}, we subjectively picked the $g$-interval $[1.4,2.4]$, which by the eye seems to give a nice powerlaw close to a straight line. To introduce \emph{some} way to estimate the systematic error, we selected a few subintervals contained in the basic interval, and fitted the parameters $\Delta_\epsilon$, $g_c$ for each such
subinterval.\footnote{In the future, the fit procedure could be refined by taking into account the value of $\mathcal{E}_2-\mathcal{E}_0$ at $g=g_c$.} We obtained  $g_c = 3.04(15)$ and $\Delta_\epsilon = 1.06(13)$. This value of $\Delta_\eps$ is compatible with the two-dimensional Ising model value $\Delta_\epsilon = 1$, giving us confidence that the procedure is sensible. To improve the estimate of $g_c$, we fix $\Delta_\eps$ to this theoretically known value and redo the fit. We then get $\bar g_c =  2.97(3)$.

The above error estimate may be too optimistic, because we completely ignored the error in $m_\phys$ induced by truncation effects. We have also performed the fit taking the $L=10$, $\Emax=20$ `renormalized subleading' results as central values, and the difference $\sigma$ between these central values and the `renormalized' results without subleading correction as the error (we consider the two-sided error $\pm\sigma$). Following this procedure and doing the fit in the $[1.4,2.4]$ interval we obtained $\bar g_c =  2.97(14)$. This is our final, conservative, estimate.

\begin{figure}[t!]
\begin{center}
  \includegraphics[scale=0.7]{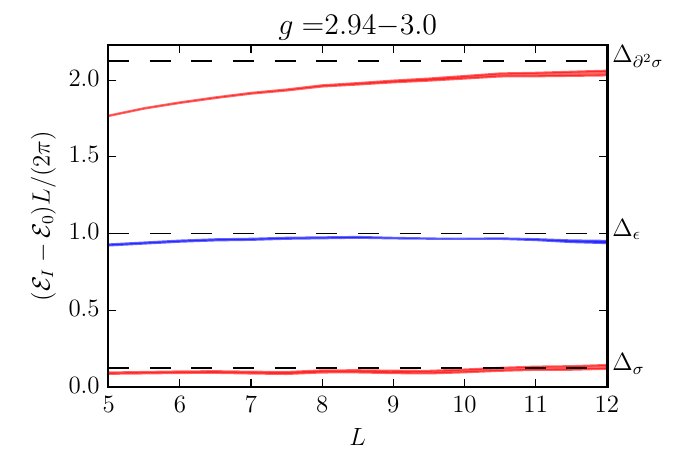}
  \end{center}
\caption{Comparison with the CFT spectrum; see the text.}
\label{fig:plotVsL_specNorm}
\end{figure}

We now perform another comparison with the theoretically known CFT operator dimensions. Namely, for $g=g_c$ the excitations $\mathcal{E}_I-\mathcal{E}_0$ should go as
\beq
\mathcal{E}_I-\mathcal{E}_0 \sim \frac{2\pi}{L}\Delta_I\,,
\label{eq:CFTdim}
\eeq
where $\Delta_I$ are the CFT dimensions. This asymptotics should be valid for $L\gg 1$ where the theory has flown sufficiently close to the IR fixed point. To check this, in figure \ref{fig:plotVsL_specNorm} we plot the three lowest excitation energies multiplied by $L/(2 \pi)$. 

In this figure, we consider $L=5\ldots 12$ and vary the quartic coupling within our `optimistic' uncertainty range around the fixed point, $g = 2.94\ldots 3.0$. We have to vary the UV cutoff $\Emax$ as a function of $L$ in order to have a manageable number of basis elements in the low energy truncated Hilbert space $\mathcal{H}_l$. So $\Emax$ decreases from $33$ at $L=5$ to $18$ at $L=12$, while the truncated Hilbert space dimension stays for each $L$ around 10000 - 15000 per $\bZ_2$ sector. To avoid clutter, we show only the `renormalized subleading' results (but see figure \ref{fig:plotVsL_g=2.97} below, where the results without subleading corrections are also shown). 

As figure \ref{fig:plotVsL_specNorm} demonstrates, \reef{eq:CFTdim} \emph{is} approximately obeyed at large $L$, provided that we use the 2D Ising operator dimensions $\Delta_\sigma = 1/8$, $\Delta_\epsilon = 1$, $\Delta_{\del^2 \sigma} = 2+1/8$, where this latter operator is a scalar descendant of $\sigma$.

\subsection{$L$ dependence}
\label{sec:spectrumvsL}

We will now present several plots which show explicitly how the spectrum of the theory varies for increasing $L$ while keeping $g$ fixed. These plots are analogous to figure \ref{fig:phi4_multL}, but the information is presented somewhat differently.

\noindent{\bf $\bZ_2$-unbroken phase}

\noindent Let us look first at the $\bZ_2$-unbroken phase. We fix $g =1$, which is at the outer border or the perturbativity range (see appendix \ref{sec:pert}). Figure \ref{fig:plotVsL_g=1} shows then the vacuum energy density $\calE_0/L$ and the spectrum, for $L=5\ldots 12$.

\begin{figure}[h!]
\begin{center}
      \includegraphics[scale=0.8]{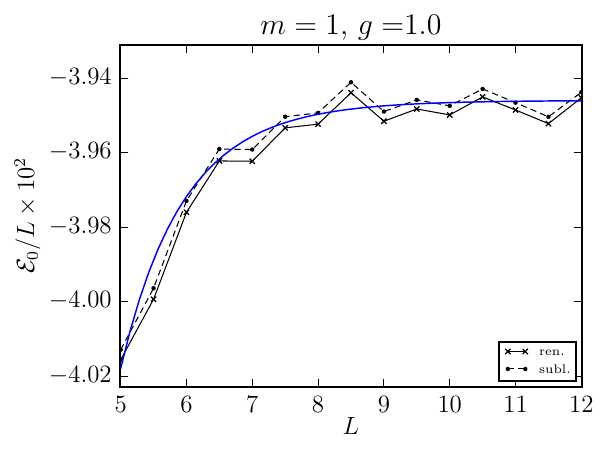} 
       \includegraphics[scale=0.8]{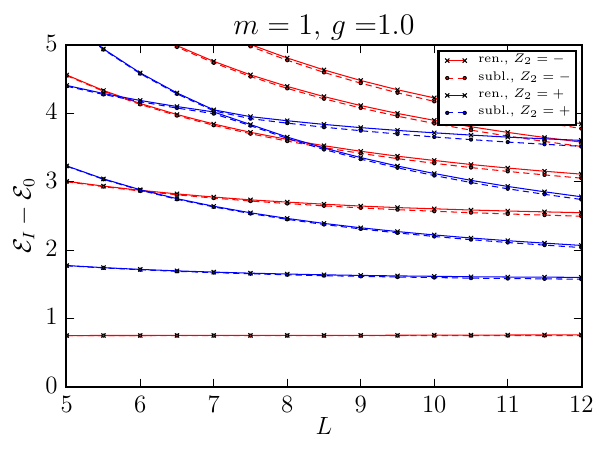}
  \end{center}
\caption{The vacuum energy density and the excitation spectrum for $g=1$, as a function of $L$.}
\label{fig:plotVsL_g=1}
\end{figure}

In the left plot we see that the vacuum energy density tends to a constant value. We don't worry too much about the fluctuations around the limit which happen for some values of $L$, like an upward fluctuation for $L=8.5$ or a downward fluctuation for $L=11.5$. These fluctuations are due to the fact that in our renormalization procedure we neglected the discreteness of the distribution $M(E)$, replacing it by a continuous approximation. As figure \ref{fig:me} shows, this approximation is meant to work only on average. The sharpness of the cutoff $E\le \Emax$ disrupts the validity of the approximation, and must be behind the above fluctuations. In the future it will be important to find a way to work around these fluctuations. One way would be to consider a cutoff which is not totally sharp.\footnote{Ref.~\cite{Hogervorst:2014rta}, section 6.4 and appendix D, describes a method which for conformal bases used in that work allowed to perform renormalization taking into account the discreteness of the sequence $M(E)$. It's not clear if that method extends to the massive Fock space bases used here.}

Ignoring for the time being the fluctuations, let us discuss the approach of the vacuum energy density to its infinite volume limit.
As is well known, in a massive phase the rate of this approach is exponentially fast and is given by:\begin{align}
\calE_0(L)/L &= \Lambda -\frac{m_\phys}{\pi L} K_1(m_\phys L)+O(e^{-2 m_\phys L})\nn\\
&\approx \Lambda -\left(\frac{m_\phys} {2 \pi L^3}\right)^{1/2} e^{- m_\phys L}\qquad (L\gg 1/m_\phys)\,.
\label{eq:largeL}
\end{align}
This formula can be derived by considering the partition function of the theory on a torus $S^1_L\times S^1_{L'}$ where $L$ and $L'$ are the lengths of the circles. The $\calE_0(L)$ is extracted by considering the limit $L'\gg L$, and so it's natural to treat $L'$ as space and $L$ as the inverse temperature. The condition $L\gg 1/m_\phys$ means that we are interested in low temperatures. The deviation of the free energy can then be described in terms of thermodynamics of a gas of particles of mass $m_\phys$. This type of arguments is standard in the thermodynamic Bethe ansatz calculations in integrable theories, in which case also the subleading terms in \reef{eq:largeL} can be determined; see 
e.g.~\cite{Zamolodchikov:1989cf}, Eq.~(3.13). However, the leading term that we show is more general. It does not require integrability nor knowing anything about how the particles interact - we can treat them as free in this computation. In fact 
\reef{eq:largeL} can be also determined by taking the large $L$ limit of the free scalar Casimir energy \reef{eq:E0L} with $m\to m_\phys$.

The blue curve in the left plot is the fit of our numerical data by Eq.~\reef{eq:largeL} with $m_\phys$ fixed to the value determined from the numerical spectrum (see below). We see that the rate of the approach to the infinite $L$ limit is reasonably well described by the theoretically predicted dependence.\footnote{Since $m_{\rm \ph} < m$, the effect we are observing here is formally dominant with respect to the exponentially suppressed $E_0(L)$ and $z(L)$ corrections, which were omitted in section \ref{sec:probl_meth1}. Still, the hierarchy $m/m_{\rm \ph}$ is not very large, and a more careful comparison may be warranted in the future, taking also those corrections into account.}

The accompanying right plot shows the spectrum of excitations above the vacuum. Observe the remarkably small difference between the two renormalization procedures (we use this difference as an idea about the error of the method).
The first excited state in the odd sector should for large $L$ approach the infinite-volume physical mass $m_\phys$. It shows hardly any variation with $L$ in the shown range, which is consistent with the rate of approach being exponentially fast in $m_\phys L$ \cite{Luscher:1985dn}. We extract $m_\phys=0.751(1)$.

The second excited state, which belongs to the even sector, for large $L$ asymptotes to 1.47(4) which within error bars coincides with $2 m_\phys$. This state corresponds to having two particles at rest. Notice that we do not observe any states in the energy range between $m_\phys$ and $2 m_\phys$. Such states would be interpreted as two-particle bound states. As is well known, the $\phi^4$ interaction is perturbatively repulsive, so we do not expect bound states at weak coupling. Moreover it is known rigorously that two-particle bound states are absent everywhere below the phase transition; see \cite{Glimm:1987ng}, section 17.2. What we observe here is consistent with these results.

Notice that the `two-particles at rest' state approaching $2m_\ph$, as well as the `three-particles at rest' state going to $3 m_\ph$, show a much larger variation with $L$ compared to the one-particle state. That this variation is \emph{not} exponentially suppressed is a consequence of particle-particle interactions. Since the interactions are short-ranged, their effect is expected to go like the inverse volume, $1/L$ \cite{Luscher:1986pf}. It should be possible to use this effect to extract information about the two-particle $S$-matrix.\footnote{Such analyses are standard in the TCSA approach to $d=2$ RG flows; see \cite{Yurov:1991my,Lencses:2014tba} for the first and a recent example.}

For small $g$, it is easy to calculate these corrections explicitly using the first-order perturbation theory for the Hamiltonain \reef{eq:hamiltonian}. For the two-particle and three-particle states at rest we get\footnote{These formulas are valid for a fixed finite $L$ and $g\ll \pi^2 m/L$. In this limit the splittings between different states with the same number of particles are sufficiently large so that we can neglect their mixing. In the opposite limit one should apply quasi-degenerate perturbation theory.}  
\beq
\mathcal{E}_2 = 2 m + \frac{3 g}{L m^2}+ O(g^2)\,,\qquad \mathcal{E}_3 = 3 m + \frac{9 g}{L m^2}+ O(g^2)\,.
\eeq
The positiveness of the $O(g)$ corrections explains the ``bumps'' at small coupling in the corresponding curves in figure \ref{fig:phi4} (the first $\bZ_2$-even and the second $\bZ_2$-odd states).
 
The even state just above the one asymptoting to $2m_\phys$ should be identified as corresponding to two particles moving in the opposite directions on the circle with one unit of momentum each. Using the one-particle dispersion relation, the energy of this state should be roughly $2\times (m_\phys^2 + (2 \pi /L)^2)^{1/2}$ plus the corrections due to the particle interactions in finite volume. Because of the $2\pi$ prefactor, the dispersion relation corrections are significant even at the maximal values of $L$ that we are considering; they seem to explain most of the difference between the first two even states. At larger $L$, we expect the particle interaction corrections to take over, since their strength decreases only as $1/L$.

Our final comment about the $g=1$ spectrum plot concerns the pattern of level crossings. In a non-integrable quantum field theory, we do not expect energy levels of the same symmetry to cross when varying the volume. In fact, the absence or presence of level crossings can be used as an empirical check of integrability (see \cite{Brandino:2010sv} for a related recent discussion). Since the $\phi^4$ theory is, for all we know, non-integrable, levels with the same $\bZ_2$ quantum number should not cross. Most levels in figure \ref{fig:plotVsL_g=1} do not cross trivially because they never come close each other. However, there is one interesting ``avoided" crossing: the third and fourth $\bZ_2=+$ levels head for a collision around $L=7$ but then repel. Many more such avoidances are present in the higher energy spectrum (not shown in figure \ref{fig:plotVsL_g=1}).

\noindent {\bf The critical point}

\begin{figure}[h!]
\begin{center}
  \includegraphics[scale=0.8]{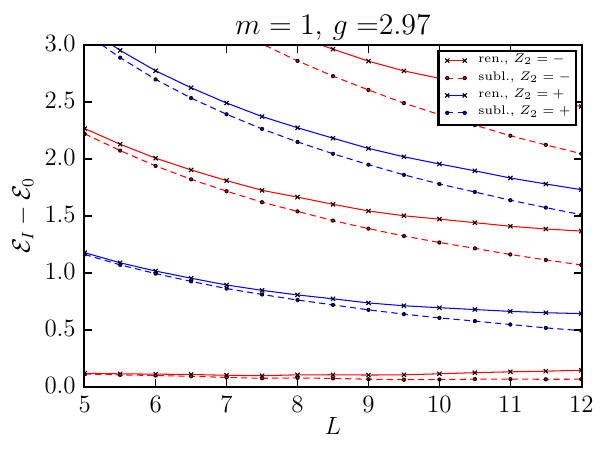}
  \includegraphics[scale=0.8]{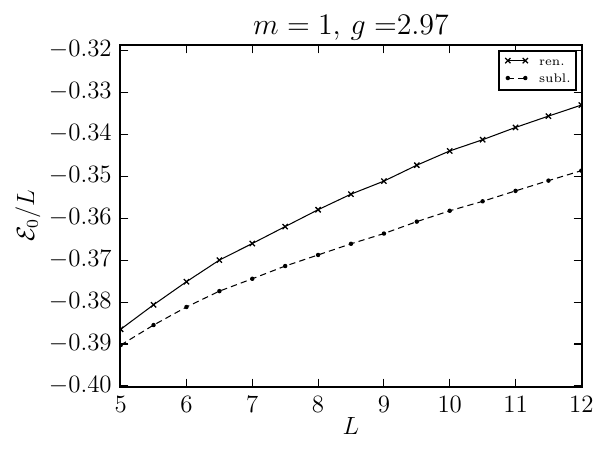} 
  \end{center}
\caption{Same as in figure \ref{fig:plotVsL_g=1}, but for $g=2.97$.}
\label{fig:plotVsL_g=2.97}
\end{figure}

\noindent In figure \ref{fig:plotVsL_g=2.97} we show analogous plots for the neighborhood of the critical point. We fix $g = 2.97$, i.e.~the central value for our $g_c$ estimate. One drastic change compared to figure \ref{fig:plotVsL_g=1} is that the energy differences $\mathcal{E}_I -\mathcal{E}_0$ (plotted on the left) no longer tend to constants but scale as $1/L$, as expected for a CFT. This is the same plot as in figure \ref{fig:plotVsL_specNorm}, except that here we do not multiply by $L/2\pi$, and we show results for both renormalization methods, to get an idea of possible error bars. Evidently, even if $g$ is not exactly equal to the critical coupling, the mass gap is sufficiently small so that it is not visible for the values of $L$ shown in this plot.

On the right we show the vacuum energy density, which, as expected, seems to approach a constant. However, the uncertainty, measured by whether or not we include the subleading corrections, remains significant.
Theoretically, the asymptotics of approach to the limit should be $-\pi c /(6 L^2)$, where $c=1/2$ is the central charge of the critical point. Instead, we see something like a $1/L$ approach. Clearly, one should work to reduce the truncation errors before the agreement is achieved.

It should be remarked that the vacuum energy is always subject to larger errors than the spectrum of excitations. This is related to the fact that the unit operator, whose coefficient shifts the vacuum energy, is the most relevant operator of the theory, and gets the largest renormalization when the states above $\Emax$ are integrated out. However, whichever uncertainty in the coefficient of the unit operator cancels when we compute the spectrum of excitations.

\subsection{$\Emax$ dependence}
\label{sec:Emaxdep}

To get a better feel for the convergence of our method, and to demonstrate the difference between the three procedures explained in section \ref{sec:proc}, we will present here plots of the spectrum and vacuum energy as a function of $\Emax$, while keeping the other parameters fixed. 

\begin{figure}[h!]
\begin{center}
  \includegraphics[scale=0.9]{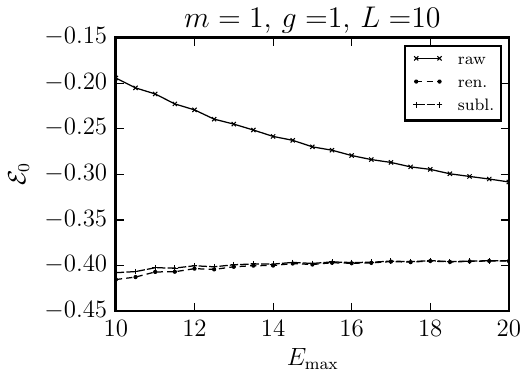} 
    \includegraphics[scale=0.9]{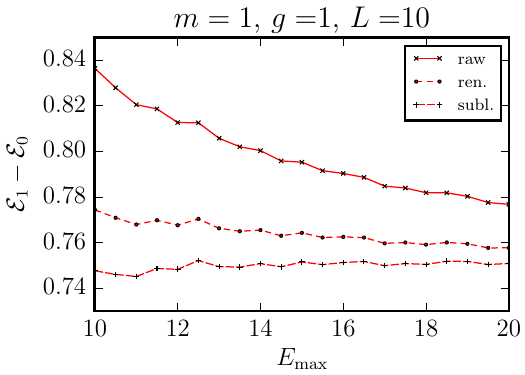}
  \end{center}
\caption{Variation with $\Emax$ and the effect of renormalization corrections for $g=1$.}
\label{fig:plotVsEmax_g=1}
\end{figure}

So, figure \ref{fig:plotVsEmax_g=1} shows the results for $g=1$, $L=10$, with $\Emax$ varying from $10$ to $20$. On the left we see that the renormalization dramatically improves the convergence of the vacuum energy with respect to the raw results, while the subsequent subleading correction is very small. The plot on the right refers to the first excited level ($i=1$). In this case we see that the further improvement due to the subleading correction is non-negligible. There are small oscillations due to discretization effects, as already discussed in section \ref{sec:spectrumvsL}. The higher excitations, not shown in the plot, show a similar pattern of convergence.

\begin{figure}[h!]
\begin{center}
  \includegraphics[scale=0.8]{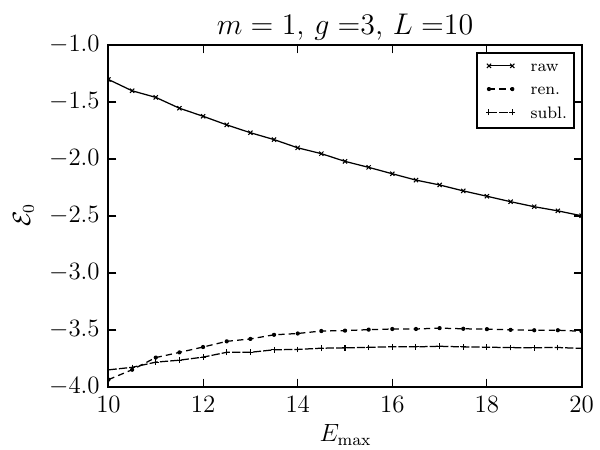} 
    \includegraphics[scale=0.8]{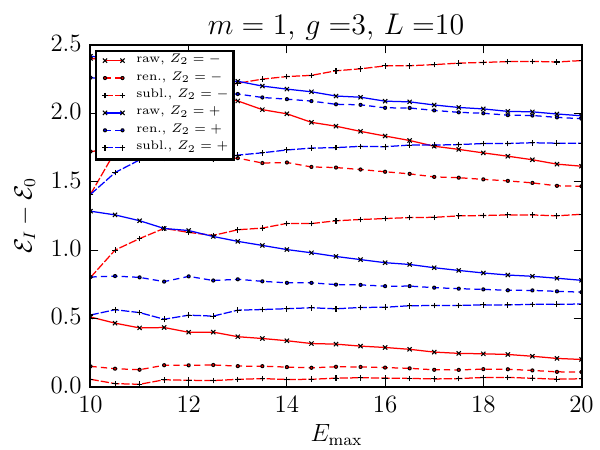}
  \end{center}
\caption{Same as in figure \ref{fig:plotVsEmax_g=1} but for $g=3$.}
\label{fig:plotVsEmax_g=3}
\end{figure}

Figure \ref{fig:plotVsEmax_g=1} shows the same plots for $g=3$. Once again the improvements due to renormalization are evident. For a change, here we show more states in the spectrum of excitations.

\subsection{Comparison to the TCSA methods}
\label{sec:tcsa}

As already mentioned, Ref.~\cite{Hogervorst:2014rta} recently studied the $\phi^4$ theory in $d=2.5$ dimensions using the TCSA method. The results of that study, and in particular the phase diagram of the theory, turned out to be quite similar to the one we found here; see \cite{Hogervorst:2014rta}, section 7. The TCSA uses the basis of conformal operators of the free massless scalar field theory, which via the state-operator correspondence is the same as the basis of states of this theory put on the sphere $S^{d-1}$. In the TCSA, both the $\phi^2$ and $\phi^4$ perturbations are included into the $V$ part of the Hamiltonian. This should be contrasted with our current method, where $\phi^2$ is included into $H_0$. We will mention here just one advantage and one complication of working with the conformal basis and treating all potential terms as a perturbation. The advantage is that the Hamiltonian matrix $H_{ij}$ for a general sphere radius $R$ is related to the $R=1$ matrix via a simple rescaling. The complication is that the conformal basis is not orthonormal, requiring introduction of a Gram matrix or dealing with an eigenvalue problem which is not symmetric.

There were several reasons why \cite{Hogervorst:2014rta} considered $d=2.5$. First of all, the main point of that paper was to show that the TCSA works in $d>2$. Second, there were technical reasons to postpone the physically more interesting case $d=3$ to the future. The final reason is that, at least naively, conformal basis does not work in $d=2$, because the scalar field dimension becomes zero, rendering the spectrum dense and numerical treatment impossible.


In spite of this basic difficulty, a recent paper \cite{Coser:2014lla} proposed a way to use the conformal basis in $d=2$ dimensions. The idea of this work is to compactify the free scalar boson on a circle of a finite length $2\pi/\beta$. Compactification renders the CFT spectrum discrete, and one hopes that for a sufficiently small $\beta$ compactification effects will be negligible. It's important to realize that the procedure of \cite{Coser:2014lla} modifies the quantum mechanical dynamics only for the zero mode, while all higher oscillator modes don't feel 
it.\footnote{For example, it would be wrong to think of their procedure as considering the scalar boson in a quartic potential cut off at the boundaries of the interval $[-\pi/\beta,\pi/\beta]$ and periodically extended to the whole real line. Such a periodized potential would not even give a UV-complete theory, because of the spikes at the cutoff points.}

On the conceptual level, the difference between our paper and \cite{Coser:2014lla} lies in the choice of the trial wavefunction basis for the oscillators modes. They choose periodic plane waves on a circle of radius $2\pi/\beta$ for the zero mode, and harmonic oscillator wavefunctions
of frequency $2\pi |n|/L$ for the modes with $|n|>0$. We instead choose harmonic oscillator wavefunctions of frequency $\sqrt{m^2+(2\pi n/L)^2}$ for all modes. Of course the technique for evaluating the matrix elements is also different, since we use prosaic ladder operators, while they are able to use the Kac-Moody algebra acting in the free scalar boson CFT.

Apart from $\beta$ which we will not discuss further, the basic parameters used in \cite{Coser:2014lla} to parametrize the phase diagram are: the length of the spatial circle $R$, which is the same as our $L$, and the quadratic and quartic couplings $G_{2,4}$.\footnote{These are denoted $g_{2,4}$ in \cite{Coser:2014lla}, but we capitalized to avoid the confusion with our notation in section \ref{sec:renormalization}.} The latter translate to our parameters as follows:\footnote{The factor $2\pi$ in the quartic arises from the difference of the $\phi$ normalization. The extra term in $m^2$ appears from the difference in implementing the normal ordering prescription, see their Eq.~(65) and the discussion in \cite{Rychkov:2015vap}.} 
\begin{gather}
g=2\pi  G_4\,,\\
m^2=G_2+\frac{6g}{\pi}\log [e^\gamma mL/(4\pi)]\,.
\end{gather}
In the $\bZ_2$-preserving phase, their strongest coupled point had $G_2=0.01$ and $G_4=8\times 10^{-5}$, which gives $\bar g=g/m^2\approx 0.05$. From our perspective, this is an extremely weakly coupled case, where even ordinary perturbation theory would be largely adequate. 

It appears that in the $\bZ_2$-preserving phase our trial wavefunction basis for the zero mode is more efficient than that of \cite{Coser:2014lla}, since it consists of wavefunctions peaked at $\phi_0=0$, as opposed to being evenly spread over a long interval.
We hasten to add however that the main goal of \cite{Coser:2014lla} was to study the $\bZ_2$-broken phase in the regime of negative $m^2$, something that we have not even attempted in this paper. In our forthcoming work \cite{Rychkov:2015vap}, dedicated to the $\bZ_2$-broken phase, careful choice of the wavefunction basis for the zero mode will also play an important role.

\section{Comparison with prior work}
\label{sec:priorwork}

The $\phi^4$ theory in two dimensions has been previously studied, in the strongly coupled region, with a variety of techniques. Table~\ref{table:gc} summarizes the predictions for the critical coupling. Here we only mention the methods which, at least in principle, allow for a systematic improvement of the results, leaving out simple-minded variational studies. Many of these papers normalize the quartic coupling as $\lambda/4!$; we translate all results to our normalization.

The clear trend in the table is that the critical coupling estimate seems to increase with time. The first two studies are rather old and do not assign an uncertainty to their results. The next result (DMRG) has the smallest claimed error, but as we will see below there are strong reasons to believe that it is grossly underestimated. The stated uncertainty of the two remaining predictions is also significantly smaller than ours. 
Their central values are below our result, although consistent with it at a $2\sigma$ level if we use the conservative error estimate. 
As we will discuss in section \ref{sec:MC}, this slight discrepancy may be due to a subtlety in implementing the matching to a continuum limit in their procedures.

\begin{table}[h!]
\begin{center}
\begin{tabular}{l|l|l}
Method & $\bar{g}_c$  & Year, ref. \\ \hline
DLCQ &  1.38 & 1988 \cite{Harindranath:1988zt} \\ 
QSE diagonalization &  2.5 & 2000 \cite{Lee:2000ac} \\ 
DMRG & 2.4954(4) & 2004 \cite{Sugihara:2004qr} \\ 
Lattice Monte Carlo &  $2.70^{+0.025}_{-0.013}$ & 2009 \cite{Schaich:2009jk} \\ 
Uniform matrix product states  &  2.766(5) & 2013 \cite{Milsted:2013rxa}\\ 
Renormalized Hamiltonian truncation &  2.97(14) & This work   \\ \hline
\end{tabular}
\caption{Estimates of $\bar g_c$ from various techniques.}
\label{table:gc}
\end{center}
\end{table}

We will now review the methods in Table~\ref{table:gc}, following the chronological order.
\subsection{DLCQ}
\label{sec:DLCQ}
In \cite{Harindranath:1987db,Harindranath:1988zt}, the $\phi^4$ theory was studied using the Discretized Light Cone Quantization (DLCQ). This is a Hamiltonian truncation method in which the theory is quantized in the light-cone coordinates $x^\pm=t\pm x$, using $x^-$ as `space' and $x^+$ as `time'. The Hilbert space consists of states of several particles all moving in the $x^+$ direction, and having a fixed total momentum $P^+$. This method was much touted in the past because of the apparent reduction in the number of states (since only particles moving in one direction are needed), and the simplicity of the vacuum structure, which in perturbation theory coincides with the free theory vacuum. In practical computations, one discretizes (hence \emph{Discretized} LCQ) the momentum fraction of constituent particles with a step $1/K$. This is sometimes presented as a result of compactifying the $x^-$ direction on a circle of length $2\pi K$.

Refs.~\cite{Harindranath:1987db,Harindranath:1988zt} used DLCQ to compute the physical particle mass as a function of $g$, observing that it goes to zero for a certain critical value of $g_c$. They find $\bar g_c\approx 1.83$ for $K=16$ \cite{Harindranath:1987db}, and later report an even smaller value $\bar g_c\approx 1.38$ based on extrapolating the $K\le 20$ results to $K=\infty$ \cite{Harindranath:1988zt}. These results are in a stark disagreement with the more recent calculations by other techniques in Table~\ref{table:gc}. A careful repetition of these old studies is called for. It is known that DLCQ calculations are subject to severe $1/K$ truncation effects \cite{vandeSande:1996mc}, which may be the source of the discrepancy.

We would like to mention here a recent proposal to avoid the $P^+$ discretization altogether, and instead truncate the light-cone Hilbert space by using a carefully constructed orthonormal basis of multi-particle wavefunctions. This alternative approach may be the future of the light-cone quantization. It already proved very promising in the study of 2d gauge theories \cite{Katz:2013qua,Katz:2014uoa}, but was not yet applied to the $\phi^4$ theory (see \cite{Chabysheva:2014rra} for the preparatory work).

As a final comment on the light-cone quantization, we note that the method is bound to become more complicated in the $\bZ_2$-broken phase, possibly requiring a scan of the zero mode $\langle \phi \rangle$ to find the true vacuum.

\subsection{QSE diagonalization}
 
Ref.~\cite{Lee:2000ac} (see also \cite{Lee:2000xna,Salwen:2002dx,Windoloski}) studied the $\phi^4$ theory using the Hamiltonian truncation in the same basic setup as ours, calling it ``modal field theory". However, the implementation details are quite different. They use a \emph{quasi-sparse eigenvector} (QSE) method, which reduces the Hilbert space dimension by throwing out the Fock states whose contributions to the physical eigenstate one is studying are small. In a later work \cite{Lee:2000xna} they developed a \emph{stochastic error correction} (SEC) method, which corrects for the resulting truncation. While the idea is similar to our renormalization, there are some differences. One difference is that their method is perturbative, unlike our basic equation \reef{eq:ex1} which is all-order in $\Delta H$. Another difference is that SEC computes infinite sums involved in the definition of $\Delta H$ via Monte Carlo sampling, while we found an analytic approximation for this correction term. 

\begin{figure}[htbp]
\begin{center}
\includegraphics[scale=0.8]{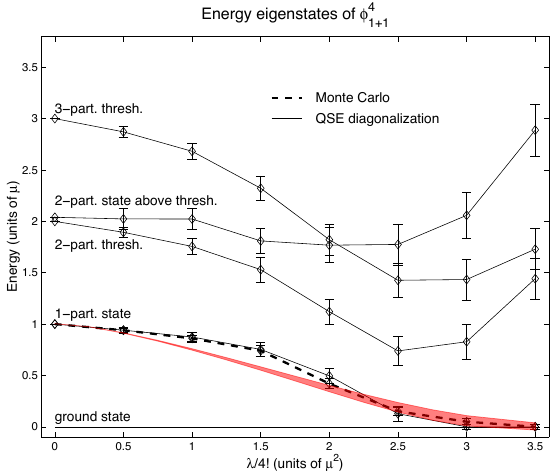}
\end{center}
\caption{Finite volume spectrum of the $\phi^4$ theory on a circle of length $L=10\pi m^{-1}$ (plot taken from \cite{Lee:2000ac}). In our notation $\lambda/4!=g$, $\mu=m$. Black solid lines with error bars---the results of QSE with 250 states.  Black dashed line---the results of a lattice Monte Carlo simulation. On their plot we overlay our results for the lowest $\bZ_2$-odd state on a circle of a smaller length $L=10m^{-1}$ (red band). The central value and the width of the red band are the same as in the conservative method of determining $\bar g_c$ in section \ref{sec:fixed}.}
\label{lee}
\end{figure}

In figure \ref{lee} we show their results for the finite volume spectrum \cite{Lee:2000ac}. These results are based on QSE with 250 states (no SEC). Using this plot, Ref.~\cite{Lee:2000ac} estimated the critical coupling as $\bar{g}_c \approx 2.5$. On the same plot we overlay our results for the lowest $\bZ_2$-odd state from figure \ref{fig:phi4}. Our predictions for the physical mass are in disagreement with \cite{Lee:2000ac} in the range $\bar g \lesssim 2$, where the truncation errors due to finite $\Emax$ are small. Notice that even though our results refer to a smaller value of $L$ than \cite{Lee:2000ac}, this cannot explain the differences, since the finite volume effects for the one-particle state are negligible in this range of $\bar g$ (see figure \ref{fig:plotVsL_g=1}). One possible explanation is that the momentum cutoff $k_{\max}=4m$ used in \cite{Lee:2000ac} is not sufficiently high to describe the continuum limit. In any case, it is this disagreement which is ultimately responsible for the difference in our estimates of $\bar g_c$. 

The QSE method of \cite{Lee:2000xna} looks somewhat similar in spirit to the Numerical RG (NRG) method recently employed in the context of TCSA \cite{Konik:2007cb,Brandino:2010sv}. At the same time, the latter method seems to us more flexible and systematic. It would be interesting to apply the NRG method to the $\phi^4$ theory and see if it can help resolve the above discrepancy.

\subsection{DMRG}
\label{sec:DMRG}

Ref.~\cite{Sugihara:2004qr} studied the $\phi^4$ theory using the Density Matrix Renormalization Group (DMRG) \cite{WhitePRB}. As a starting point of this approach, the $x$-direction is discretized with a spacing $a$, while time is kept continuous. The Hamiltonian describing such a discretized theory is
\beq
\label{eq:dmrg}
H=\sum_x \frac{1}{2a}\pi_x^2 + \frac{1}{2a}(\phi_x-\phi_{x+a})^2+\frac {m^2 a}{2} \phi_x^2 + ga \,\phi_x^4\,,
\eeq
where $\phi_x$ are the field variables on each lattice site and $\pi_x$ are the corresponding canonical momenta. The Hilbert space on each site is infinite, unlike in the more standard DMRG applications. Ref.~\cite{Sugihara:2004qr} truncates this Hilbert space to $N=10$ first harmonic oscillator states. The finite-system version of the DMRG algorithm \cite{WhitePRB} is used, truncating to $M=10$ most dominant density matrix eigenstates. This corresponds to the superblock Hamiltonian dimension $M^2 N=1000$. 

The critical value of the coupling is obtained approaching the critical point from inside of the $\bZ_2$-broken region, and studying how the vacuum expectation value $\langle \phi \rangle$ approaches zero in this limit. The quoted value has an extremely small uncertainty: $\bar g_c=2.4954(4)$. However, careful reading of the paper leaves us unconvinced that all sources of systematic error were properly taken into account. First, no attempt is made at extrapolating to $M=\infty$, while Figure 4 of \cite{Sugihara:2004qr} shows clearly that convergence in $M$ is slow and the results for $M=10$ have not yet stabilized. Second, the value of $\bar g_c$ is determined in Figure 7 of \cite{Sugihara:2004qr} by fitting a straight line through two points. 

Finally, we believe that the matching to the continuum limit should have been done more carefully. 
In the units $m^2=1$, the smallest physical lattice spacing in \cite{Sugihara:2004qr} is $a\approx 0.1$.\footnote{This is found from $\bar{g}_c a^2=\tilde \lambda/4!$ where their smallest $\tilde \lambda=0.6$.} 
This is factor 3 larger than the spacing used in the lattice Monte Carlo study \cite{Schaich:2009jk} discussed in section \ref{sec:MC} below. Since Ref.~\cite{Sugihara:2004qr} used the simplest nearest-neighbor discretization of the $x$-derivative, the matching procedure will likely be plagued by the same basic problem as the one we will explain in section \ref{sec:MC}.

\subsection{Lattice Monte Carlo}
\label{sec:MC}

In \cite{Schaich:2009jk} (see \cite{Loinaz:1997az} for earlier work) the critical coupling of the $\phi^4$ theory was determined by Monte Carlo (MC) simulations on the two-dimensional square lattice. They find $\bar g_c=2.7^{+0.025}_{-0.01}$, somewhat below our prediction. This $2\sigma$ discrepancy is not necessarily a reason to worry,
as it may go away with further development of our method. In addition, it appears that the MC computation is subject to a subtle systematic error which was not discussed in \cite{Schaich:2009jk}. This error is particularly troubling because similar errors likely affect, to varying degree, all techniques involving the discretization of space, including also the DMRG and MPS methods discussed in sections \ref{sec:DMRG} and \ref{sec:uMPS}. Below we will review the lattice computation and explain this potential error.

Ref.~\cite{Schaich:2009jk} simulated the lattice action (the subscript $\lat$ stands for ``lattice")
\beq
S_\lat=a^2\sum _x \half \sum_{\mu=1,2}a^{-2}(\phi_{x+a e_\mu}-\phi_x)^2+\half m_\lat^2 \phi_x^2+g_\lat\NO{\phi^4_x}.\label{eq:Slat}
\eeq
Here $a$ is the lattice spacing. The normal ordering on the lattice is defined by subtracting a loop of the lattice propagator (BZ = the Brillouin zone $|p_\mu|\le \pi/a$):
\begin{gather}
\NO{\phi^4_x}=\phi_x^4-\phi_x^2\int_{\rm BZ} \frac{dp}{(2\pi)^2} G_\lat(p)\,,\label{eq:phi4lat}\\
G_\lat(p)=\left\{4a^{-2}[\sin^2 (p_1 a/2)+\sin^2 (p_2 a/2)]+m_\lat^2\right\}^{-1}\,. \label{eq:Glat}
\end{gather}
So operationally, \reef{eq:phi4lat} is plugged into \reef{eq:Slat} and the resulting action is MC-simulated.

In the normalization in which $m_{\lat}=1$, Ref.~\cite{Schaich:2009jk} explored the range of lattice spacings $a=0.3$ - $0.03$.\footnote{See their Table II. The value of $a$ is computed from $\hat \mu_c^2 =m^2_{\lat} a^2$.} Their lattices had up to 1024$\times$1024 sites, which corresponds to a sufficiently large physical volume varying from $L\approx 300$ for $a=0.3$ to $L\approx 30$ for $a=0.03$. Depending on $a$, the critical quartic coupling was found to vary from $g_\lat\approx 2.55$ to $2.7$. Their final answer for $g_c$ was obtained by fitting and extrapolating to $a=0$.

The systematic error that we have in mind concerns the matching between the lattice and the continuum. Naively, the lattice theory \reef{eq:Slat} seems to go to the continuum limit theory as $a\to0$, with $m_\lat$ and $g_\lat$ turning into $m$ and $g$. However, let us try to establish this correspondence more carefully. 

\begin{figure}[h!]
\begin{center}
\includegraphics[scale=1]{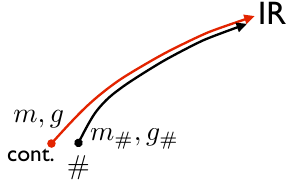}
  \end{center}
\caption{The lattice and the continuum RG flows should agree in the IR. See the text.}
\label{fig:two-flows}
\end{figure}

In figure  \ref{fig:two-flows} we show, schematically, two RG flows: the lattice flow specified by the couplings $m_\lat,g_\lat$ and the continuum flow specified by $m,g$. The latter couplings have to be found so that the flows become the same at large distances. We can check if this is the case computing some observables at intermediate distances, when the flows are still perturbative.\footnote{We are focussing on the case when the coupling $g$ is strong, which is relevant for the critical point. The case of small $g$ is simpler, as the matching can be performed at $p\lesssim m$.} If a sufficient number of observable agree at intermediate distances, the two flows have converged and will stay the same also at larger distances. In the language of effective field theory, this would be an example of perturbative matching (see e.g.~\cite{Rothstein:2003mp}).\footnote{In this discussion we ignore another complication arising from the fact that the two-dimensional $\phi^4$ theory has infinitely many additional relevant couplings beyond $m^2$ and $g$, since all powers of $\phi$ are relevant. Strictly speaking establishing correspondence between the lattice and the continuum may require turning on these extra couplings.}

At what distance scale should we do the matching? First of all, to match the continuum theory, the lattice theory should at the very least become approximately rotationally invariant. The leading deviation from rotation invariance comes from the lattice propagator \reef{eq:Glat}, which at small momenta behaves as
\beq
G_\lat^{-1}(p)=p^2+m_\lat^2-\frac 1{12} (p_1^4+p_2^4)a^2+\ldots
\eeq
To ensure that this is approximately rotationally invariant, we must have $p^2 \ll a^{-2}$.

On the other hand, the matching momentum cannot be too small since the theory is then strongly coupled.
The smallest allowed matching momentum can be computed by considering the diagrams which give a correction to the quartic coupling. For momenta $p\gg m$ these diagrams are, omitting logarithmic factors,
\beq
\vcenter{\hbox{\includegraphics[trim= 1cm 0.5cm 1cm 0.5cm, clip=true, scale=0.45]{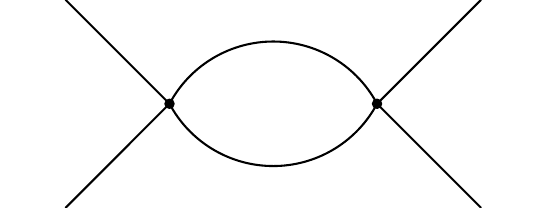}}} + {\rm permutations} \sim g^2/p^2,
 \eeq
which becomes comparable to the coupling $g$ itself for $p^2=O(g)$. Putting the two constraints together, we conclude that the matching must be done at momenta $p$ such that
\beq
g \ll p^2 \ll a^{-2}\,.
\eeq
Now, to match the mass, we have to consider the correction to the propagator, which in the considered region of momenta behaves like
\beq
\vcenter{\hbox{\includegraphics[trim= 0cm 1.7cm 0cm 1.7cm, clip=true, scale=0.6]{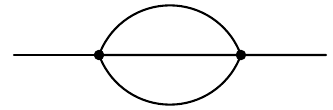}}} \,\,\, \sim g^2/p^2[1+O(p^2a^2)]
\eeq
where the terms dependent on $a^2$ indicate the schematic dependence of the correction on the lattice spacing. This suggests that 
\beq
m^2=m_\lat^2+O(g^2 a^2)\,.
\eeq
However, such a conclusion would be on shaky grounds. The problem is that at the lowest allowed momenta $p^2\sim g$ the correction to the propagator due to the rotation invariance breaking has the same parametric order of magnitude, $g^2 a^2$, as the putative mass matching correction. 

The above discussion suggests that the chosen form of the lattice discretization prevents performing a controlled matching between the lattice and the continuum theory, because the matching corrections from loop diagrams cannot be cleanly disentangled from the rotation invariance breaking effects in the propagator. This may seem unusual to a lattice practitioner. However, the theory we are considering is a bit unusual, having a coupling constant of dimension exactly 2. 

We consider it possible that this problem contributes to the mismatch between the lattice determination of $g_c$ and our results. Our discussion also suggests the recipe to remedy the problem: one should redo the lattice simulation using an improved actions, in which the leading $O(p^2a^2)$ effect of rotation symmetry breaking is absent due to judiciously chosen next-to-nearest interaction terms \cite{Symanzik:1983dc}. In such a setup the matching can be done, and the correspondence between $m_\lat,g_\lat$ and $m,g$ can be established rigorously.

\subsection{Uniform matrix product states}
\label{sec:uMPS}

This method was applied to the $\phi^4$ theory in \cite{Milsted:2013rxa}. The starting point of this approach is the discretized Hamiltonian \reef{eq:dmrg}. The lowest energy states are searched for in a finite variational subspace of the full Hilbert space, consisting of the so-called matrix product states (MPS), whose precise definition can be found in \cite{Milsted:2013rxa}. The MPS states are parametrized by a 3-tensor of size $d\times D \times D$. Here, $d$ represents the size of the truncated Hilbert space per lattice site, while $D$ is a parameter which bounds the degree of entanglement of the ground state across different lattice sites. The variational states are found by minimizing the energy through an imaginary-time evolution algorithm. The physical predictions are recovered in the limit $d,D \to \infty$, $a \to 0$.

As is well known, the MPS methods are essentially equivalent to DMRG (see e.g.~\cite{Schollwock}). Comparing with the DMRG study in section \ref{sec:uMPS}, $d$ and $D$ should be identified with $N$ and $M$. Ref.~\cite{Milsted:2013rxa} uses $d=16$ and $D$ up to 128, commenting that $N=M=10$ used in \cite{Sugihara:2004qr} are not sufficient. They observe that an insufficiently large $D$ shifts the critical point to lower $\bar g_c$, and provide a physical explanation for this effect. They do two measurements of $\bar g_c$, both approaching the critical point from above, one using $\langle \phi\rangle$ and another from the lowest excitation energy. Since their two measurements differ at a $3\sigma$ level, the value cited in Table \ref{table:gc} was obtained by expanding the error bars to include both of them.

In the units $m^2=1$, the minimal value of the lattice spacing in \cite{Milsted:2013rxa} is $a\approx 0.04$, about the same as in \cite{Schaich:2009jk}. This study is thus subject to the same worries about the matching to the continuum limit as the ones brought up in section \ref{sec:MC}. 

\section{Discussion}
\label{sec:conclusions}

In this work we revisited one of the simplest realizations of the ``exact diagonalization'' methods, as opposed to standard lattice Monte Carlo methods, and shown that it can be used effectively as a numerical tool to extract non-perturbative predictions about a quantum field theory. The numerical setup is relatively simple, and the error coming from the UV regulator can be reduced by adding analytically computed correction terms to the Hamiltonian. 

Our choice of the model to study here---the two-dimensional $\phi^4$ theory---was dictated by several considerations:
\begin{itemize}
\item the model is not supersymmetric nor integrable, hence not amenable to analytical methods, apart from perturbation theory at small coupling\,;
\item the model has been studied in the past by a variety of numerical techniques, allowing for a fair comparison of the results and of the implementation difficulties\,; 
\item the model is literally \emph{the} textbook example of a quantum field theory. In fact we hope that our exercise also has a considerable pedagogical value, helping to bridge the conceptual gap between perturbative and non-perturbative QFT questions.
\end{itemize}
However we stress that the idea of the paper is completely general, and it should be possible to apply similar techniques to any quantum field theory.

In this exploratory work we did not push particularly hard on the numerical side of the calculations---it takes a few single-core days on a desktop to reproduce all the plots in this paper. Our analytical calculations of the renormalization coefficients can and will be advanced, further improving the accuracy. The current state of the method allowed us to compute the low-energy spectrum in the $\bZ_2$-invariant phase with a reasonable accuracy, and to observe qualitatively the change to the $\bZ_2$-broken phase at strong coupling. Our estimate for the critical coupling is in a slight disagreement with the existing results. As discussed in section \ref{sec:MC}, this may be partly due to a technical subtlety in the lattice regularization. It would be interesting to resolve this tension in future work.

Comparisons with other Hamiltonian truncation techniques, such as TCSA or light-cone quantization, are scattered throughout the paper (see sections \ref{sec:compMat}, \ref{sec:tcsa}, \ref{sec:DLCQ}). At this point in history we don't want to be religious about which one of these methods is most promising---all have to be explored without prejudice to see which one gives more accurate predictions, depending perhaps on the problem under consideration. One of the main challenges for all these techniques is their application to higher dimension, where the truncated Hilbert space for a given UV cutoff is larger, while the interesting interaction terms are less relevant, resulting in more significant truncation errors. In the TCSA context, these issues recently started being addressed in \cite{Hogervorst:2014rta}. Another challenge is the application to gauge theories. Here the light-cone quantization seems to have gained an upper hand, at least in $d=2$, thanks to the extremely efficient conformal bases recently proposed in \cite{Katz:2013qua,Katz:2014uoa}.

The grand question at stake is---shall we live to see the computation of the proton mass becoming accessible to every theorist, or will it forever remain in the realm of dedicated collaborations wielding supercomputers?
Currently computations of the low-energy QCD spectrum with $2+1$ dynamical quark flavors with a few percent accuracy take about one supercomputer-year, roughly equivalent to a $100,000$ single-core-years. 

\section*{Acknowledgements}
We thank Daniele Dorigoni, Matthijs Hogervorst, Robert Konik, Giuseppe Mussardo, Agostino Patella, David Simmons-Duffin and Balt van Rees for the useful discussions, Dean Lee for the permission to use the plot in figure \ref{lee}, and Mark Windoloski and Dean Lee for providing a copy of \cite{Windoloski}. This research was partly supported by the National Centre of Competence in Research SwissMAP, funded by the Swiss National Science Foundation. The work of L.V. is supported by the Swiss National Science Foundation under grant 200020-150060.

\appendix
\section{Speeding up the Hamiltonian matrix computation}
\label{sec:speed}

In our computations, most time is spent in matrix diagonalization. Still, matrix evaluation should also be organized efficiently. Here we list some tricks useful to speed it up. These tricks are realized in our {\tt python} code, included with the {\tt arXiv} submission.

\noindent {\bf Diagonal/offdiagonal decomposition} 
\nopagebreak

\noindent Let's split $H$ into three parts:
\beq
\label{eq:Hsplit}
H=H_{\rm diag}+H_{\rm offdiag}+H^\dagger_{\rm offdiag}
\eeq
where $H_{\rm diag/offdiag}$ have only diagonal/offdiagonal matrix elements. $H_{\rm diag}$ includes $H_0$ and the terms in $V$ of the form\footnote{Here and below $\{x_1,x_2,\ldots\}$ denotes an unordered set.}
\beq
a^\dagger_{k_1}a^\dagger_{k_2}a_{k_3}a_{k_4},\quad \{k_1,k_2\} = \{k_3,k_4\}.
\eeq 
The rest of the terms in $V$ get assigned to $H_{\rm offdiag}$ and $H_{\rm offdiag}^\dagger$. Only the matrix elements of $H_{\rm offdiag}$ need to be evaluated, while those of $H_{\rm offdiag}^\dagger$ are obtained by transposition. We include into $H_{\rm offdiag}$ the $a^\dagger a^\dagger a^\dagger a^\dagger$, $a^\dagger a^\dagger a^\dagger a$ terms in $V$, 
as well as the operators
\beq
\label{eq:ccaa}
a^\dagger_{k_1} a^\dagger_{k_2} a_{k_3} a_{k_4},\quad \{k_1,k_2\} \ne \{k_3,k_4\}\,,
\eeq  
satisfying the following lexicographic ordering condition:\footnote{It's not hard to see that ${\rm sort}(|k_1|,|k_2|)= {\rm sort}(|k_3|,|k_4|)$ is impossible given $\{k_1,k_2\} \ne \{k_3,k_4\}$ and $k_1+k_2=k_3+k_4$. So any operator \reef{eq:ccaa} gets assigned either to $H_{\rm offdiag}$ or to $H^\dagger_{\rm offdiag}$.}
\beq
\label{eq:lex}
{\rm sort}(|k_1|,|k_2|)\prec {\rm sort}(|k_3|,|k_4|)
\eeq
Notice that this condition depends only on the absolute values of momenta, hence it is $\bP$-invariant. This ensures that all three terms in the decomposition \reef{eq:Hsplit} are separately $\bP$-invariant. This will be important below, when we describe our method to evaluate the matrix elements.

\noindent{\bf Keeping track of the energy}
\nopagebreak

\noindent Each elementary operator $\calO\in V$, a product of ladder operators, increases/decreases energy of any basis vector it acts upon by a fixed amount $\Delta E_\calO$. Since we will be working in the space of low-energy states $\calH_l$ of energies $0\le E\le \Emax$, we can drop from $V$ all operators for which 
\beq
|\Delta E_\calO|> \Emax\,.
\eeq
Moreover, when acting on a basis state $\ket{\psi}$ the result is guaranteed to be zero in $\calH_l$ unless
\beq
0 \le E(\psi)+\Delta E_\calO\le \Emax\,.
\eeq

\noindent{\bf Combinatorial factors for oscillator ordering}
\nopagebreak

\noindent To reduce the number of elementary operators in $V$, it's worth ordering them. We have
\beq
\sum_{k_1,k_2,k_3,k_4}  a_{k_1}a_{k_2}a_{k_3}a_{k_4} = \sum_{k_1\le k_2\le k_3 \le k_4} f_4(k_1,k_2,k_3,k_4) a_{k_1}a_{k_2}a_{k_3}a_{k_4}
\eeq
where the symmetry factor
\beq
f_4(a\le b\le c\le d)=\begin{cases}
24& a<b<c<d\,, \\
12& a=b<c<d\text{ or }a<b=c<d\text{ or }a<b<c=d\,,\\
6& a=b<c=d\,, \\
4 & a=b=c<d\text{ or }a<b=c=d\,,\\
1 & a=b=c=d\,.
\end{cases}
\eeq

\noindent{\bf $\bP$-conservation}
\nopagebreak

\noindent In this paper we work in the Hilbert space of $P=0$ states of energies $E\le \Emax$.
Internally we represent a state $\ket{\psi}$, see \reef{eq:state}, as a sequence of occupation numbers $Z_n$ for each momentum mode:
\beq
\label{eq:intern}
\ket{\psi} \leftrightarrow [Z_n:-\nmax\le n\le \nmax]\,,
\eeq
where $\nmax$ is the maximal possible mode number for the given $L$ and $\Emax$.

The matrix $H_{ij}$ is then computed as follows. The diagonal part from $H_0$ is trivial so we do not discuss it. For the rest, we take a particular state $\ket{\psi_j}$ and act on it with elementary operators $\calO\in V$, one by one. Each operator gives one particular state $\ket{\psi_i}$ times a numerical coefficient. We accumulate this coefficient in the matrix element $H_{ij}$. Thus the matrix is generated column by column. As discussed above, we can do this computation for $H_{\rm offdiag}$ and get $H^\dagger_{\rm offdiag}$ by transposition. We generate the matrix separately in each of the $\bZ_2=\pm$ sectors.

The computation we just discussed produces the matrix $H$ in the full Hilbert space of $P=0$, $E\le \Emax$ states. However, in this paper we are interested in the $\bP=+1$ subspace of this space. The basis of this subspace consists of symmetrized linear combinations \reef{eq:symm} of the basic $P=0$ Fock states. In principle, the matrix in the $\bP=+1$ subspace could be obtained once the full matrix is computed, but this is wasteful. We will now describe a method which generates the matrix in the $\bP=+1$ subspace directly.

\def\sym{{\rm sym}}
When we store the symmetrized state $\ket{\psi^{\rm sym}}$ internally, we only store $\ket{\psi}$. If $\ket{\psi}\ne \bP\ket{\psi}$, then we keep only one of these two vectors (no matter which one), since they give rise to the same $\ket{\psi^{\rm sym}}$. 

We have to compute the matrix with respect to the symmerized basis, which we will call $S_{ij}$:
\beq
H\ket{\psi_j^{\sym}}=S_{ij}\ket{\psi_{i}^\sym}\,.
\eeq
Consider also the matrix $H_{ij}$ with respect to the Fock basis, whose computation was discussed above. Let's split it into three pieces:
\beq
H\ket{\psi_i}=H^a_{ji}\ket{\psi_j}+H^b_{ki}\ket{\psi_k}+H^c_{ki}\bP\ket{\psi_k}\,,
\eeq
where the index $j$ runs over $\bP$-invariant $\ket{\psi_j}$, and the rest of the Fock basis is split into $\ket{\psi_k}$'s and $\bP\ket{\psi_k}$'s. Since $[\bP,H]=0$, we have
\beq
\label{eq:pinvused}
H \bP\ket{\psi_i}=\bP(H \ket{\psi_i})= H^a_{ji}\ket{\psi_j}+H^b_{ki}\bP\ket{\psi_k}+H^c_{ki}\ket{{\psi_k}}\,,
\eeq
and finally
\begin{align}
H \ket{\psi^\sym_i}=\beta(\psi_i)(H \ket{\psi_i}+ H \bP\ket{\psi_i})&= \beta(\psi_i) [2H^a_{ji}\ket{\psi_j}+
(H^b_{ki}+H^c_{ki})(\ket{\psi_k}+\bP\ket{\psi_k})]\nn
\\
&=\beta(\psi_i)  [2H^a_{ji}\ket{\psi^\sym_j}+
\sqrt{2}(H^b_{ki}+H^c_{ki})\ket{\psi^\sym_k}]
\end{align}
From here we obtain a recipe for an economic way to compute $S_{ji}$. 
Namely, we compute $H\ket{\psi_i}$ and accumulate the coefficients $2H^a_{ji}$ and $\sqrt{2}(H^b_{ki}+H^c_{ki})$,
and then multiply by $\beta(\psi_i)$. 

Notice that we used the $\bP$-invariance of the Hamiltonian in the first step of \reef{eq:pinvused}. When this method is combined with splitting $H$ into the diagonal/offdiagonal parts, it's important that every part be $\bP$-invariant by itself.
As mentioned above, condition \reef{eq:lex} ensures this.

%
%

\section{Perturbation theory checks}
\label{sec:pert}

{\red Some statements in this appendix are wrong, see Note Added below.}

We computed the first two perturbative corrections to the ground state energy density $\Lambda$ and the physical particle mass for the $\phi^4$ theory defined by the action~\reef{eq:action}: 
\begin{gather}
\Lambda/m^2 = -
\frac{21 \zeta(3)}{16 \pi^3}\bar{g}^2+ 0.0416485  \bar{g}^3+\ldots,
\label{eq:Lambdapert}\\
\Delta m^2/m^2 \equiv (m_\phys^2 - m^2)/m^2=  -\frac{3}{2} \bar{g}^2+ 2.86460(20) \bar{g}^3+\ldots
\label{eq:dm2pert}
\end{gather}
($\bar{g}\equiv g/m^2$). Recall that $\Lambda$ at $g=0$ is set to zero. Because the interaction is normal ordered the $O(\bar{g})$ contributions are absent. The $O(\bar g^3)$ coefficients are numerical with a shown number of significant digits and an error estimate if needed.\footnote{It's likely that exact expressions for these coefficients can be found, but since this is not the focus of our work, we have not invested the effort.} The size of the coefficients suggests that the series are perturbative for $\bar g\lesssim 1$.

The coefficients were obtained by numerical integration of Feynman diagrams. It is much easier to perform this integration in the coordinate space, since the propagator \reef{eq:prop} is exponentially decreasing at large distances, and also because parallel lines in multiloop diagrams correspond in the $x$-space to trivially raising the propagator to a power. For example, the $O(g^3)$ correction to $\Delta m^2$ comes from the diagram
\beq
\includegraphics[trim= 0cm 2cm 0cm 2cm, clip=true, scale=0.8]{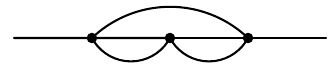}
\eeq
evaluated at the (Euclidean) external momentum $p^2=-m^2$. In the $x$-space this gives the integral (we omit the combinatorial factors)
\beq
\int d^2 x \int d^2 y\, e^{ip.x}\, G(|x-y|)^2\, G(|y|)^2\, G(|x|)\,.
\eeq
We pick $p=(i m,0)$, introduce the polar coordinates and evaluate the integral via Monte Carlo.

\begin{figure}[h!]
\begin{center}
\includegraphics[scale=0.8]{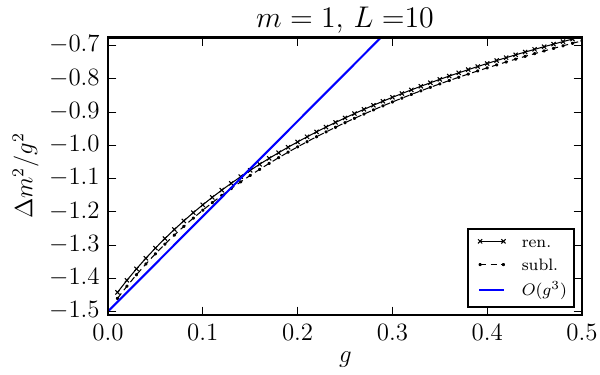}
\includegraphics[scale=0.8]{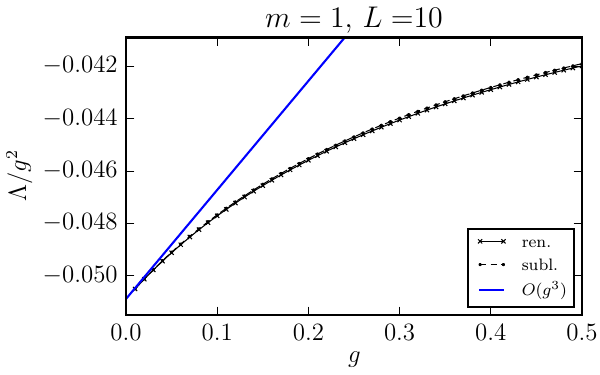} 
\end{center}
\caption{Comparing perturbative and numerical predictions; see the text.}
\label{fig:phi4pert}
\end{figure}

In figure \ref{fig:phi4pert} we compare the above perturbative results with the numerical spectra obtained with our method for $m=1$, $L=10$. Perturbative computations refer to the infinite volume, but $L=10$ is sufficiently large so that the expected exponentially small corrections should not disturb the comparison. We use the cutoff $\Emax=20$. Notice that $m_\phys$ is extracted as $\calE_1-\calE_0$, where  $\calE_1$ is the lowest $\bZ_2$-odd eigenstate, while $\Lambda$ is extracted as $\calE_0/L$. 

To facilitate the comparison, we plot $\Lambda$ and $\Delta m^2$ divided by $g^2$. The reasonably good match in the region of small $g\lesssim0.1$ shows that our numerical method agrees with both $O(\bar g^2)$ and $O(\bar g^3)$ coefficients of the perturbative expansion. At the same time, higher order corrections are clearly non-negligible---they would become comparable to the $O(\bar g^3)$ correction at $\bar g\sim0.5$. 

It should be noticed that it has been rigorously shown in the constructive field theory literature that perturbation theory in the two-dimensional $\phi^4$ theory is Borel-summable for small $\bar g$; see \cite{eckmann1975} and the discussion in \cite{Glimm:1987ng}, section 23.2. Using Lipatov-type arguments \cite{Lipatov:1977hj,Brezin:1976vw}, the asymptotic behavior of the perturbative series coefficients is predicted to be\footnote{The order of magnitude of coefficients (but not the alternating signs) were justified rigorously in \cite{Breen:1984vh}.}
\begin{gather}
(-1)^k k^k A^k\,,\label{eq:lip}\qquad A=\min \int d^2 x  \bigl(\half (\del \psi)^2+\half \psi^2  - \lambda \psi^4 \bigr) -\log \lambda\,,
\end{gather}
where one has to look for a saddle point in $\psi$ and $\lambda$ which gives the minimal $A$.
Given this asymptotics, one could hope that the Borel transform is regular for all positive $\bar g$, with a leading singularity at the negative coupling $\bar g=-A$. It is not obvious to us how this analytic structure would be compatible with the phase transition at a finite $\bar g\approx 3$.

As a side remark, we notice that the two-dimensional $\phi^4$ theory in the $\bZ_2$-symmetric phase
seems sufficiently simple so that the perturbation theory can be worked out, by a numerical integration of Feynman integrals, to a very high order. The asymptotic behavior of the coefficients can be also worked out with many subleading terms. Given that, we would like to challenge the resurgence/Borel transform community (see e.g.~\cite{Cherman:2013yfa}) to reproduce the dependence $m_\ph(\bar g)$ with a precision matching that of our method.

\begin{figure}[h!]
	\begin{center}
		\includegraphics[scale=0.8]{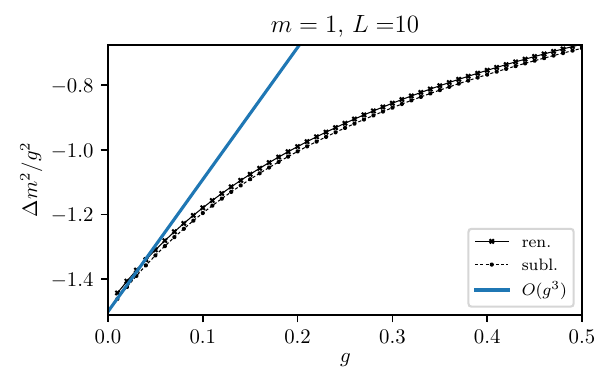}
	\end{center}
	\caption{Comparing perturbative and numerical predictions for the mass after correcting $O(g^3)$ perturbative coefficients.}
	\label{fig:phi4pert-corr}
\end{figure}

{\bf Note added} (Aug 2018) The $O(g^3)$ coefficient in \reef{eq:dm2pert} is wrong as we forgot to include another diagram which contributes this coefficient. With both diagrams included, the value of the coefficient changes to $\frac{9}{\pi} \text{(our diagram)} + \frac{63 \zeta(3)}{2 \pi^3} \text{(missed diagram)}\approx 4.086$. Analytic values of both diagrams were computed in \cite{Serone:2018gjo} which reacted to the challenge stated in the previous paragraph. With the coefficient corrected, the perturbative prediction becomes tangent to the numerical one at small $g$; see Fig.~\ref{fig:phi4pert-corr}. We thank Marco Serone, Gabriele Spada and Giovanni Villadoro for their magnificent paper, and for informing us about our mistake.

\footnotesize

\bibliography{phi4-Biblio}
\bibliographystyle{utphys}

\end{document}